\documentclass[runningheads]{llncs}

\usepackage{fontawesome}  
\usepackage{amsmath}
\usepackage{amssymb}

\usepackage{multirow}
\usepackage[table,xcdraw]{xcolor}
\usepackage{graphicx}
\usepackage{textcomp}
\usepackage{subfig}
\usepackage[utf8x]{inputenc}
\usepackage{csquotes}
\usepackage{flexisym}
\newcommand{\ignore}[1]{}
\usepackage[hidelinks]{hyperref}
\usepackage[font=footnotesize]{caption}

\usepackage[ruled,vlined,linesnumbered]{algorithm2e}

\usepackage[bottom]{footmisc}
\usepackage{wrapfig}
\usepackage{nicefrac}
\usepackage{bm}
\usepackage{xcolor}

\newcommand{\activityUniverse}{\mathcal{A}}
\newcommand{\resourceUniverse}{\mathcal{R}}
\newcommand{\caseUniverse}{\mathcal{C}}
\newcommand{\sensitiveUniverse}{\mathcal{S}}
\newcommand{\processUniverse}{\mathcal{P}}
\newcommand{\timeUniverse}{\mathcal{T}}
\newcommand{\otherUniverse}{\mathcal{D}}

\newcommand{\multiset}{\mathcal{B}}
\newcommand{\mainEventUniverse}{\mathcal{E}}
\newcommand{\tlkc}{\mathit{TLKC}}
\newcommand{\perspectiveUniverse}{\mathcal{PS}}
\newcommand{\orcid}[1]{
	\href{https://orcid.org/#1}{\includegraphics[scale=0.4]{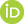}}
}

\usepackage{mathtools}

\begin{document}
	\title{Group-Based Privacy Preservation Techniques for Process Mining}
	\titlerunning{Group-Based Privacy Preservation Techniques for Process Mining}
	%
	%
	%
	\author{Majid Rafiei\orcid{0000-0001-7161-6927}\textsuperscript{\href{mailto:majid.rafiei@pads.rwth-aachen.de}{\faEnvelopeO}} \and
		Wil M.P. van der Aalst\orcid{0000-0002-0955-6940}}
	\authorrunning{Majid Rafiei and Wil M.P. van der Aalst}
	%
	\institute{Chair of Process and Data Science, RWTH Aachen University, Aachen, Germany \\
	}
	\maketitle              

	\begin{abstract}
		Process mining techniques help to improve processes using event data. Such data are widely available in information systems. However, they often contain highly sensitive information.
		For example, healthcare information systems record event data that can be utilized by process mining techniques to improve the treatment process, reduce patient's waiting times, improve resource productivity, etc. However, the recorded event data include highly sensitive information related to treatment activities.
		Responsible process mining should provide insights about the underlying processes, yet, at the same time, it should not reveal sensitive information. 
		In this paper, we discuss the challenges regarding directly applying existing well-known group-based privacy preservation techniques, e.g., $k$-anonymity, $l$-diversity, etc, to event data. 
		We provide formal definitions of attack models and introduce an effective \textit{group-based privacy preservation technique} for process mining.
		Our technique covers the main perspectives of process mining including \textit{control-flow}, \textit{time}, \textit{case}, and \textit{organizational} perspectives. The proposed technique provides interpretable and adjustable parameters to handle different privacy aspects.
		We employ real-life event data and evaluate both data utility and result utility to show the effectiveness of the privacy preservation technique. We also compare this approach with other group-based approaches for privacy-preserving event data publishing.

		
	\end{abstract}
	\section{Introduction}\label{sec:introduction}
	
	Process mining employs event data to discover, analyze, and improve the real processes \cite{van2016process}. 
	Indeed, it provides fact-based insights into the actual processes using event logs.
	There are many algorithms and techniques in the field of process mining. 
	However, the three basic types of process mining are (1) \textit{process discovery}, where the goal is to learn real process models from event logs, (2) \textit{conformance checking}, where the aim is to find commonalities and discordances between a process model and an event log, and (3) \textit{process re-engineering} (\textit{enhancement}), where the aim is to extend or improve a process model using different aspects of the available data.
	
	An event log is a collection of events where each event is described by its attributes \cite{van2016process}. 
	The typical attributes required for the main process mining algorithms are \textit{case identifier}, \textit{activity}, \textit{timestamp}, and \textit{resource}.
	The \textit{case identifier} refers to the entity that the event belongs to, the \textit{activity} refers to the activity associated with the event, the \textit{timestamp} is the time that the event occurred, and the \textit{resource} is the activity performer.
	In the human-centered processes, case identifiers refer to persons. For example, in a patient treatment process, the case identifiers refer to the patients whose data are recorded. 
	Moreover, the \textit{resource} attribute often refers to the persons performing activities, e.g., in the healthcare context, the resources refer to the doctors or nurses performing activities for the patients. 
	The event attributes are not limited to the above-mentioned ones, and an event may also carry other case-related attributes, so-called case attributes, e.g., \textit{age}, \textit{salary}, \textit{disease}, etc, which could be considered as sensitive person-specific information. Table~\ref{sample_evenlog} shows a sample event log.
	
	Orthogonal to the three mentioned types of process mining, different perspectives are also defined including \textit{control-flow},  \textit{organizational},  \textit{case}, and \textit{time} perspective \cite{van2016process}. 
	The \textit{control-flow perspective} focuses on activities and their order, which are often utilized by \textit{process discovery} and \textit{conformance checking} techniques. 
	The \textit{organizational perspective} focuses on resources and their relations, which are exploited by \textit{social network discovery} techniques. 
	The \textit{case perspective} is focused on case-related attributes, and the \textit{time perspective} is concerned with the time-related information, which can be used for \textit{performance and bottleneck analyses}.
	
	With respect to the main attributes of events, two different perspectives for privacy in process mining can be considered in the human-centered processes; \textit{resource perspective} and \textit{case perspective}. The \textit{resource perspective} focuses on the privacy rights of the individuals performing activities, and the \textit{case perspective} concerns the privacy rights of the individuals whose data are recorded and analyzed. Depending on the context, the relative importance of these perspectives may differ. However, often the \textit{case perspective} is more critical for privacy than the \textit{resource perspective}. For example, in the healthcare context, activity performers could be publicly available. However, what happens for a specific patient and her/his personal information should be kept private. In this paper, we are focused on the \textit{case perspective}. 
	In principle, when event logs explicitly or implicitly include personal data, \textit{privacy concerns} appear which should be taken into account according to regulations such as the European General Data Protection Regulation (GDPR) \cite{voss2016european}.
	
	\begin{figure}[t]
		\centering
		\includegraphics[width=0.99\textwidth]{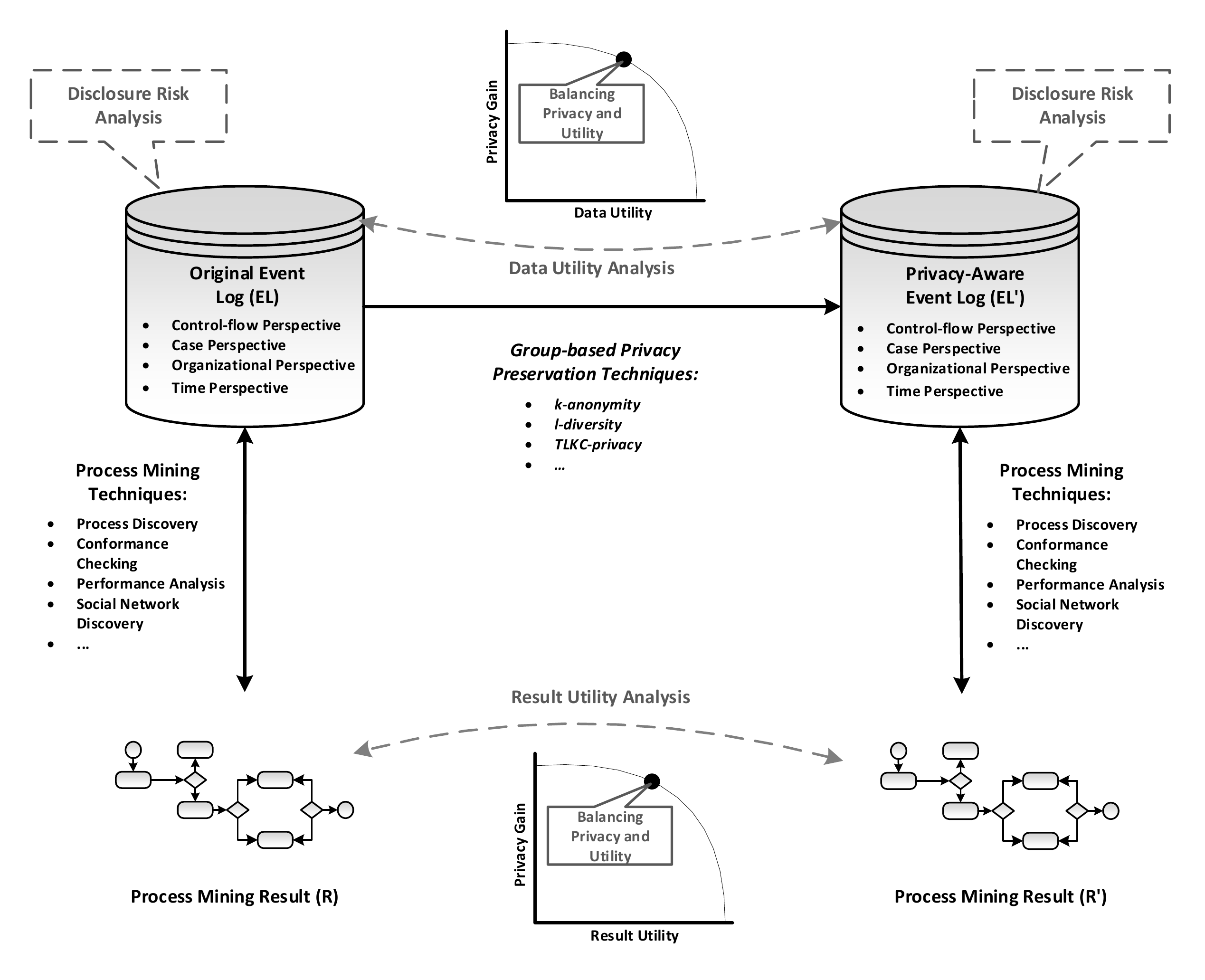}
		\caption{The general overview of privacy-related activities in process mining. Privacy preservation techniques are applied to event logs to mitigate disclosure risks. The data and result utility analyses are used to evaluate the effectiveness of the techniques where the goal is to balance utility loss and privacy gain.}\label{fig:overview}
	\end{figure} 

	In this paper, we describe \textit{disclosure risks} and \textit{linkage attacks} against event logs. The attack models are formally defined based on the available event attributes.   
	We discuss the challenges regarding directly applying group-based privacy preservation techniques, e.g., $k$-anonymity \cite{kanonymity}, $l$-diversity \cite{ldiversity}, etc, to event logs.
	We extend the work described in \cite{rafieitlkc}, where the $\tlkc$-privacy is introduced as an effective group-based privacy preservation technique for process mining.
	The $\tlkc$-privacy exploits some restrictions regarding the availability of background knowledge in the real world to deal with process mining-specific challenges. This technique is focused on \textit{control-flow}, \textit{time}, and \textit{case} perspectives. 
	$\tlkc$-privacy generalizes several traditional privacy preservation techniques, such as $k$-anonymity, confidence bounding \cite{confidenceBounding}, ($\alpha$,$k$)-anonymity \cite{alpha-k}, and $l$-diversity.
	
	The extended privacy preservation technique covers all the main perspectives of process mining including \textit{control-flow}, \textit{time}, \textit{case}, and \textit{organizational} perspectives. It empowers the adjustability of the proposed technique by adding new parameters to adjust privacy guarantees and the loss of accuracy. 
	Moreover, a new utility measure is defined to tackle the drawbacks of the current approach. 
	To evaluate the extended technique, we employ real-life event logs and evaluate both \textit{data utility} and \textit{result utility}. We also compare the extended $\tlkc$-privacy with the main algorithm and other group-based approaches for privacy-preserving event data publishing.
	Our experiments show that the proposed approach maintains high data and result utility, assuming realistic types of background knowledge. 
	\autoref{fig:overview} shows a general overview of privacy-related activities in process mining which are discussed in this paper.

	The rest of the paper is organized as follows. In Section~\ref{sec:motivation}, we explain the motivation and challenges. Section~\ref{sec:prelimineries} provides preliminaries on event logs and different types of background knowledge. In Section~\ref{sec:attackmodel}, we provide formal models of the attacks. Privacy preservation techniques are discussed in Section~\ref{sec:PrivacyModel}. In Section~\ref{sec:exp}, the experiments are presented. Section~\ref{sec:relatedwork} outlines related work, and Section~\ref{sec:conclusions} concludes the paper.
	
	%
	
	\section{Motivation and Challenges}\label{sec:motivation}
	To motivate the necessity to deal with privacy issues in process mining, we describe the disclosure risks using an example in the health-care context. Consider Table~\ref{sample_evenlog} as part of an event log recorded by an information system in a hospital. 
	Note that each case has a sequence of events that are ordered based on the timestamps. This sequence of events is called a \textit{trace} which is a mandatory attribute for a case \cite{van2016process}. For example, case 1, which could be interpreted as patient 1, is first registered by employee 4, then visited by doctor 3, and at the end released from the hospital by employee 6.    
	
	Suppose that an adversary knows that a victim patient's data are in the event log (as a \textit{case}), with little information about some event attributes that belongs to the patient, the adversary is able to connect the patient to the corresponding \textit{case id}, so-called \textit{case disclosure} \cite{rafiei_quantification}. Consequently, two types of sensitive person-specific information are revealed: (1) the complete sequence of events belonging to the case, and (2) sensitive case attributes. (1) and (2) are generally called \textit{attribute disclosure}. (1) is also called \textit{trace disclosure} that is a specific type of \textit{attribute disclosure} \cite{rafiei_quantification}.
	For example, if the adversary knows that two blood tests were performed for the victim patient, the only matching case is the case with id 2. This attack is called \textit{case linkage} attack. After the case re-identification, the sensitive case attributes are disclosed, e.g., the disease of patient 2 is \textit{infection}. This is called \textit{attribute linkage} attack. Moreover, the complete sequence of events performed for patient 2 is disclosed which contains private information, e.g., the complete sequence of activities performed for the case, the resources who performed the activities for the case, or the exact timestamp of doing a specific activity for the case. We call this attack \textit{trace linkage} which is a specific type of \textit{attribute linkage} attack. 
	
	Note that the \textit{attribute linkage} attack does not necessarily need to be launched after the \textit{case linkage}, i.e., if more than one case corresponds to the adversaries knowledge while all the matching cases have the same value for the sensitive case attribute(s) or the same sequence of event attributes (e.g., the same sequence of activities), the \textit{attribute linkage}/\textit{trace linkage} could happen without a successful \textit{case linkage} attack. For example, if the adversary knows that the activity \textit{visit} has been performed by the resource \textit{doctor} 3 for a victim patient, case 1 and case 6 match this background knowledge. However, they both have the same sequence of activities and resources ($\langle (RE,E4),(VI,D3),(RL,E6) \rangle$). Consequently, the adversary realizes the complete sequence of activities and the resources who performed the activities.

	\begin{table}[tb]
		\centering
		\scriptsize
		\caption{Sample event log (each row represents an event).}\label{sample_evenlog}
		\begin{tabular}{|l|l|l|l|l|l|}
			\hline
			$\textbf{Case Id}$ & $\textbf{Activity}$             & $\textbf{Timestamp}$        & $\textbf{Resource}$  & $\textbf{Age}$ & $\textbf{Disease}$     \\ \hline
			1       & Registration (RE)    & 01.01.2019-08:30:00 & Employee 4 (E4) & 22  & Flu         \\ 
			1       & Visit (VI)            & 01.01.2019-08:45:00 & Doctor 3 (D3)   & 22  & Flu         \\ 
			2       & Registration (RE)    & 01.01.2019-08:46:00 & Employee 1 (E1) & 30  & Infection   \\ 
			3       & Registration (RE)    & 01.01.2019-08:50:00 & Employee 1 (E1) & 32  & Infection   \\ 
			4       & Registration (RE)    & 01.01.2019-08:55:00 & Employee 4 (E4) & 29  & Poisoning   \\ 
			1       & Release (RL)         & 01.01.2019-08:58:00 & Employee 6 (E6) & 22  & Flu         \\ 
			5       & Registration (RE)    & 01.01.2019-09:00:00 & Employee 1 (E1) & 35  & Cancer      \\ 
			2       & Hospitalization (HO) & 01.01.2019-09:01:00 & Employee 3 (E3) & 30  & Infection   \\ 
			6       & Registration (RE)    & 01.01.2019-09:05:00 & Employee 4 (E4) & 35  & Corona \\ 
			4       & Visit (VI)            & 01.01.2019-09:10:00 & Doctor 2 (D2)   & 29  & Poisoning   \\ 
			5       & Visit (VI)            & 01.01.2019-09:20:00 & Doctor 2 (D2)   & 35  & Cancer      \\ 
			4       & Infusion (IN)        & 01.01.2019-09:30:00 & Nurse 2 (N2)    & 29  & Poisoning   \\ 
			5       & Hospitalization (HO) & 01.01.2019-09:55:00 & Employee 6 (E6) & 35  & Cancer      \\ 
			3       & Hospitalization (HO) & 01.01.2019-10:00:00 & Employee 3 (E3) & 32  & Infection   \\ 
			2       & Blood Test (BT)      & 01.01.2019-10:02:00 & Nurse 1 (N1)    & 30  & Infection   \\ 
			5       & Blood Test (BT)      & 01.01.2019-10:10:00 & Nurse 2 (N2)    & 35  & Cancer      \\ 
			3       & Blood Test (BT)      & 01.01.2019-10:15:00 & Nurse 1 (N1)    & 32  & Infection   \\ 
			6       & Visit (VI)            & 01.01.2019-10:20:00 & Doctor 3 (D3)   & 35  & Corona \\ 
			4       & Release (RL)         & 01.01.2019-10:30:00 & Employee 6 (E6) & 29  & Poisoning   \\ 
			6       & Release (RL)         & 01.01.2019-14:20:00 & Employee 6 (E6) & 35  & Corona \\ 
			2       & Blood Test (BT)      & 01.02.2019-08:00:00 & Nurse 1 (N1)    & 30  & Infection   \\ 
			2       & Visit (VI)            & 01.02.2019-09:30:00 & Doctor 1 (D1)   & 30  & Infection   \\ 
			3       & Visit (VI)            & 01.02.2019-13:55:00 & Doctor 1 (D1)  & 32  & Infection   \\ 
			2       & Release (RL)         & 01.02.2019-14:00:00 & Employee 2 (E2) & 30  & Infection   \\ 
			3       & Release (RL)         & 01.02.2019-14:15:00 & Employee 2 (E2) & 32  & Infection   \\ 
			5       & Release (RL)         & 01.02.2019-16:00:00 & Employee 2 (E2) & 35  & Cancer      \\ \hline
		\end{tabular}
	\end{table}
	
	Several group-based privacy preservation techniques, such as $k$-anonymity \cite{kanonymity}, $l$-diversity \cite{ldiversity}, and $t$-closeness \cite{li2007t}, have been introduced to deal with similar attacks in the context of relational databases.  
	In such techniques, the data attributes are classified into four main categories including; \textit{explicit identifiers}, \textit{quasi-identifiers}, \textit{sensitive attributes}, and \textit{non-sensitive attributes}. The \textit{explicit identifiers} are the attributes that can be used to uniquely identify the data owner, e.g., national id. The \textit{quasi-identifiers} are a set of attributes that could be exploited to uniquely identify the data owner, e.g., $\{age,gender,zipcode\}$. The \textit{sensitive attributes} consist of sensitive person-specific information, e.g., disease or salary, and the \textit{non-sensitive attributes} contain all the attributes that do not fall into the previous three categories \cite{aggarwal2008privacy}. Assuming that \textit{explicit identifiers} suppressed or replaced with dummy identifiers, the group-based privacy preservation techniques aim to perturb potential linkages by generalizing the records into equivalence classes, i.e., groups of records, having the same values on the \textit{quasi-identifier}. These techniques are effective for anonymizing relational data. However, they are not easily applicable to event data due to some specific properties of event data.
	
	In process mining, the \textit{explicit identifiers} (i.e., actual case identifiers) do not need to be stored and processed, and case identifiers are often dummy identifiers, e.g., incremental IDs.
	As described in the above-mentioned examples, a trace can be considered as a \textit{quasi-identifier} and, at the same time, as a \textit{sensitive attribute}. In other words, a complete sequence of events belonging to a case, is sensitive person-specific information, at the same time, part of a trace, i.e., only some of the event attributes, can be exploited as a \textit{quasi-identifier} to launch \textit{case linkage} and/or \textit{attribute linkage} attacks. 
	
	The \textit{quasi-identifier} role of traces in process mining causes significant challenges for group-based privacy preservation techniques because of two specific properties of event data: the \textit{high variability of traces} and the typical \textit{Pareto distribution of traces}. Considering only \textit{activity} as the main event attribute in a trace, the variability of traces in an event log is high because of the following reasons: (1) there could be tens of different activities which could happen in any order, (2) one activity or a bunch of activities could happen repetitively, and (3) traces could contain any non-zero number of activities, i.e., various lengths. Note that this variability becomes even higher when events contain more attributes, e.g., resources. 
	In an event log, trace variants are often distributed similarly to the Pareto distribution, i.e., few trace variants are frequent and many trace variants are unique. Enforcing group-based privacy-preserving approaches on little-overlapping and high-dimensional space is a significant challenge, and often valuable data needs to be suppressed in order to achieve desired privacy requirements \cite{curse_dimen}. 
	

	\section{Preliminaries}\label{sec:prelimineries}
	
	In this section, we provide formal definitions for event logs and background knowledge. These formal models will be used in the remainder for describing the attack scenarios and the approach.
	
	\subsection{Event Log}\label{subsec:ELmodel}
	We first introduce some basic notations. For a given set $A$, $A^*$ is the set of all finite sequences over $A$, and $\multiset(A)$ is the set of all multisets over the set $A$. 
	For $A_1,A_2 \in \multiset(A)$, $A_1 \subseteq A_2$ if for all $a \in A$, $A_1(a) \leq A_2(a)$.
	A finite sequence over $A$ of length $n$ is a mapping $\sigma \in \{1,...,n\} \rightarrow{A}$, represented as $\sigma = \langle a_1,a_2,...,a_n \rangle$ where $a_i = \sigma(i)$ for any $1\leq i \leq n$. $|\sigma|$ denotes the length of the sequence. 
	For $\sigma_1, \sigma_2 \in A^*$, $\sigma_1 \sqsubseteq \sigma_2$ if $\sigma_1$ is a subsequence of $\sigma_2$, e.g., $\langle a,b,c,x \rangle \sqsubseteq \langle z,x,a,b,b,c,a,b,c,x \rangle$. For $\sigma \in A^*$, $\{a \in \sigma\}$ is the set of elements in $\sigma$, and $[a \in \sigma]$ is the multiset of elements in $\sigma$, e.g., $[a \in \langle x,y,z,x,y \rangle ] = [x^2,y^2,z]$. 
	For $x=(a_1,a_2,...,a_n) \in A_1 \times A_2 \times ... \times A_n$, $\pi_{A_i}(x) = a_i$ is the projection of the tuple $x$ on the element from the domain $A_i$, $1\le i \le n$. 

	\begin{definition}[Process Instance, Trace]
		\label{def:process_instance,trace}
		We define $\processUniverse = \caseUniverse \times \mainEventUniverse^* \times \sensitiveUniverse$ as the universe of all process instances.
		$\caseUniverse$ is the universe of case identifiers.
		$\mainEventUniverse=\activityUniverse \times \resourceUniverse \times \timeUniverse$ is the universe of main event attributes for process mining where $\activityUniverse$ is the universe of activities, $\resourceUniverse$ is the universe of resources, and $\timeUniverse$ is the universe of timestamps.
		$\sensitiveUniverse \subseteq \otherUniverse_1 \cup ... \cup \otherUniverse_m$ is the universe of sensitive case attributes where $\otherUniverse_1$,...,$\otherUniverse_m$ are the universes of different case attributes, e.g., disease, salary, age, etc.
		Given a process instance $p=(c,\sigma,s) \in \processUniverse$, $\sigma \in \mainEventUniverse^*$ is called the trace attribute of the case $c$. 
	\end{definition}
	
	%
	%

	\begin{definition}[Event Log]
		\label{def:simpleEL}
		Let $\processUniverse = \caseUniverse \times \mainEventUniverse^* \times \sensitiveUniverse$ be the universe of process instances. An event log is $EL \subseteq \processUniverse$ such that if $(c_1,\sigma_1,s_1) \in EL$, $(c_2,\sigma_2,s_2) \in EL$, and $c_1=c_2$, then $\sigma_1 = \sigma_2$ and $s_1=s_2$, i.e., all the case identifiers are unique. Moreover, if $p=(c,\sigma,s) \in EL$, then $\sigma \neq \langle \rangle$.   
	\end{definition}

	\begin{definition}[Perspective, Projection]
		\label{def:proj}
		Let $\processUniverse = \caseUniverse \times \mainEventUniverse^* \times \sensitiveUniverse$ be the universe of process instances. $ps \in \{ \activityUniverse,\resourceUniverse, \activityUniverse \times \resourceUniverse,\activityUniverse \times \timeUniverse,\resourceUniverse \times \timeUniverse,\activityUniverse \times \resourceUniverse \times \timeUniverse \}$ is a perspective which can be used to project traces of an event log $EL \subseteq \processUniverse$. 
		For $\sigma=\langle (a_1,r_1,t_1),...,(a_n,r_n,t_n) \rangle \in \mainEventUniverse^*$, such that there exists $(c,\sigma,s) \in EL$, $\pi_{ps}(\sigma)$ is the projection of the trace on the given perspective, e.g., for $ps = \activityUniverse \times \resourceUniverse$, $\pi_{ps}(\sigma)=\langle (a_1,r_1),...,(a_n,r_n) \rangle$ is the projection of the trace on the activities and resources. 
		We denote $\perspectiveUniverse= \{ \activityUniverse,\resourceUniverse, \activityUniverse \times \resourceUniverse,\activityUniverse \times \timeUniverse,\resourceUniverse \times \timeUniverse,\activityUniverse \times \resourceUniverse \times \timeUniverse \}$ as the universe of perspectives.
	\end{definition}
	
	\begin{definition}[Set of Activities/Resources in an Event Log]
		\label{def:AR_EL}
		Let $\processUniverse = \caseUniverse \times \mainEventUniverse^* \times \sensitiveUniverse$ be the universe of process instances, and $EL \subseteq \processUniverse$ be an event log. $A_{EL}=\{ a \in \activityUniverse \mid \exists_{(c,\sigma,s) \in EL} {a \in \pi_{\activityUniverse}(\sigma)} \}$ is the set of activities in the event log, and $R_{EL}=\{ r \in \resourceUniverse \mid \exists_{(c,\sigma,s) \in EL}{a \in \pi_{\resourceUniverse}(\sigma)} \}$ is the set of resources in the event log.
	\end{definition}

	\begin{definition}[Set of Traces/Variants in an Event Log]
		\label{def:trace_variant}
		Let $\processUniverse = \caseUniverse \times \mainEventUniverse^* \times \sensitiveUniverse$ be the universe of process instances, $EL \subseteq \processUniverse$ be an event log, and $ps \in \perspectiveUniverse$ be a perspective. $\overline{EL}_{ps}=[ \pi_{ps}(\sigma) \mid (c,\sigma,s) \in EL ]$ is the multiset of traces in the event log w.r.t. the given perspective. $\widetilde{EL}_{ps} = \{ \pi_{ps}(\sigma) \mid (c,\sigma,s) \in EL \}$ is the set of variants, i.e., unique traces, w.r.t. the given perspective, e.g., $\widetilde{EL}_{\activityUniverse}$ is the set of unique traces w.r.t. the activities.
	\end{definition}

	\begin{definition}[Directly Follows Relations]
		\label{def:DFR}
		Let $EL {\subseteq} \processUniverse$ be an event log, $ps {\in} \{ \resourceUniverse, \activityUniverse \}$ be a perspective, $\widetilde{EL}_{ps}$ be the set of variants and $\overline{EL}_{ps}$ be the multiset of traces in the event log $EL$ w.r.t. the given perspective $ps$. $DF^{EL}_{ps}{=}\{ (x,y) \in ps {\times} ps \mid x >^{EL}_{ps} y \}$ is the set of directly follows relations w.r.t. the given perspective. $x >^{EL}_{ps} y$ iff there exists a trace $\sigma {\in} \widetilde{EL}_{ps}$ and $1 {\le} i {<} |\sigma|$, s.t., $\sigma(i)=x$ and $\sigma(i{+}1)=y$. $|x >^{EL}_{ps} y| {=} \sum_{\sigma \in \widetilde{EL}_{ps}}\overline{EL}_{ps}(\sigma) {\times} |\{ 1 {\le} i {<} |\sigma| \mid \sigma(i){=}x \wedge \sigma(i{+}1){=}y \}|$ is the number of times $x$ is followed by $y$ in $EL$.
	\end{definition}

	\begin{definition}[Variant Frequency]
		\label{def:variant_freq}
		Let $\processUniverse = \caseUniverse \times \mainEventUniverse^* \times \sensitiveUniverse$ be the universe of process instances, and $EL \subseteq \processUniverse$ be an event log. Given a perspective $ps \in \perspectiveUniverse$, $freq^{EL}_{ps}: \widetilde{EL}_{ps} \rightarrow [0,1]$ is a function that retrieves the relative frequency of the variants in the event log w.r.t. the given perspective. $freq^{EL}_{ps}(\sigma) = \nicefrac{\overline{EL}_{ps}(\sigma)}{|\overline{EL}_{ps}|}$ and  $\sum_{\sigma \in \widetilde{EL}_{ps}}freq^{EL}_{ps}(\sigma)=1$.
	\end{definition}
	
	Table$~\ref{simple_evenlog}$ shows the process instance representation of the event log shown in Table$~\ref{sample_evenlog}$, where timestamps are represented as \enquote{day-hour:minute}. In this event log, \textit{disease} is the attribute which is considered as the sensitive one.

	\begin{table}[t]
		\centering
		\scriptsize
		\caption{The process instance representation of the event log Table~\ref{sample_evenlog} (each row is a process instance where timestamps are represented as \enquote{day-hour:minute}).}\label{simple_evenlog}
		\begin{tabular}{|c|l|l|}
			\hline
			$\textbf{Case Id}$ & $\textbf{Simple Trace}$                                                                                                                                                                             & $\textbf{Disease}$ \\ \hline
			1                                 & $\langle$(RE,E4,01{-}08{:}30),(VI,D3,01{-}08{:}45),(RL,E6,01{-}08{:}58)$\rangle$                                                                                                                       & Flu                               \\ \hline
			2                                 & \begin{tabular}[c]{@{}l@{}}$\langle$(RE,E1,01{-}08{:}46),(HO,E3,01{-}09{:}01),(BT,N1,01{-}10{:}02),\\ (BT,N1,02{-}08{:}00),(VI,D1,02{-}09{:}30),(RL,E2,02{-}14{:}00)$\rangle$\end{tabular} & HIV                               \\ \hline
			3                                 & \begin{tabular}[c]{@{}l@{}}$\langle$(RE,E1,01{-}08{:}50),(HO,E3,01{-}10{:}00),(BT,N1,01{-}10{:}15),\\ (VI,D1,02{-}13{:}55),(RL,E2,02{-}14{:}15)$\rangle$\end{tabular}                          & Infection                         \\ \hline
			4                                 & \begin{tabular}[c]{@{}l@{}}$\langle$(RE,E4,01{-}08{:}55),(VI,D2,01{-}09{:}10),(IN,N2,01{-}09{:}30),\\(RL,E6,01{-}10{:}30)$\rangle$\end{tabular}                                                                                              & Poisoning                         \\ \hline
			5                                 & \begin{tabular}[c]{@{}l@{}}$\langle$(RE,E1,01{-}09{:}00),(VI,D2,01{-}09{:}20),(HO,E6,01{-}09{:}55),\\ (BT,N2,01{-}10{:}10),(RL,E2,02{-}16{:}00)$\rangle$\end{tabular}                          & Cancer                            \\ \hline
			6                                 & $\langle$(RE,E4,01{-}09{:}05),(VI,D3,01{-}10{:}20),(RL,E6,01{-}14{:}20)$\rangle$                                                                                                                       & Corona                            \\ \hline
		\end{tabular}
	\end{table}

	
	\subsection{Background Knowledge}\label{subsec:bk}
	Regarding the \textit{quasi-identifier} role of traces, we consider four main types of background knowledge including \textit{set}, \textit{multiset} (\textit{mult}), \textit{sequence} (\textit{seq}), and \textit{relative time difference} (\textit{rel}). 
	Using \textit{set} as the type of background knowledge, we assume that an adversary knows a subset of some event attributes contained in the trace attribute of a victim case. In the \textit{multiset} type of background knowledge, the assumption is that an adversary knows a subset of some event attributes included in the trace attribute of a victim case as well as the frequency of the elements. In the \textit{sequence} type of background knowledge, we suppose that an adversary knows a subsequence  of some event attributes included in the trace attribute of a victim case.
	
	The exact timestamps of events in an event log impose a high risk regarding the linkage attacks such that little time-related knowledge may easily single out specific events, and consequently the case re-identification. For performance analysis in process mining, we need to have the time-related information. However, the timestamps do not necessarily need to be the actual ones. Therefore, we make all the timestamps relative as defined in \autoref{def:relativeTimestamps}. 
	
	\begin{sloppy}
		\begin{definition}[Relative Timestamps]
			\label{def:relativeTimestamps}
			Let $\sigma = \langle (a_1,t_1), (a_2,t_2),...,(a_n,t_n) \rangle$ be a trace including the time attribute, and $t_0$ be an initial timestamp. $relative(\sigma) = \langle (a_1,t'_1), (a_2,t'_2),...,(a_n,t'_n) \rangle$ is the trace with relative timestamps such that $t'_1 = t_0$ and for each $ 1<i \le n $, $t'_i = t_i - t_1 + t_0$.
		\end{definition}
	\end{sloppy}

\begin{figure}[bt]
		\centering
		\includegraphics[width=0.99\textwidth]{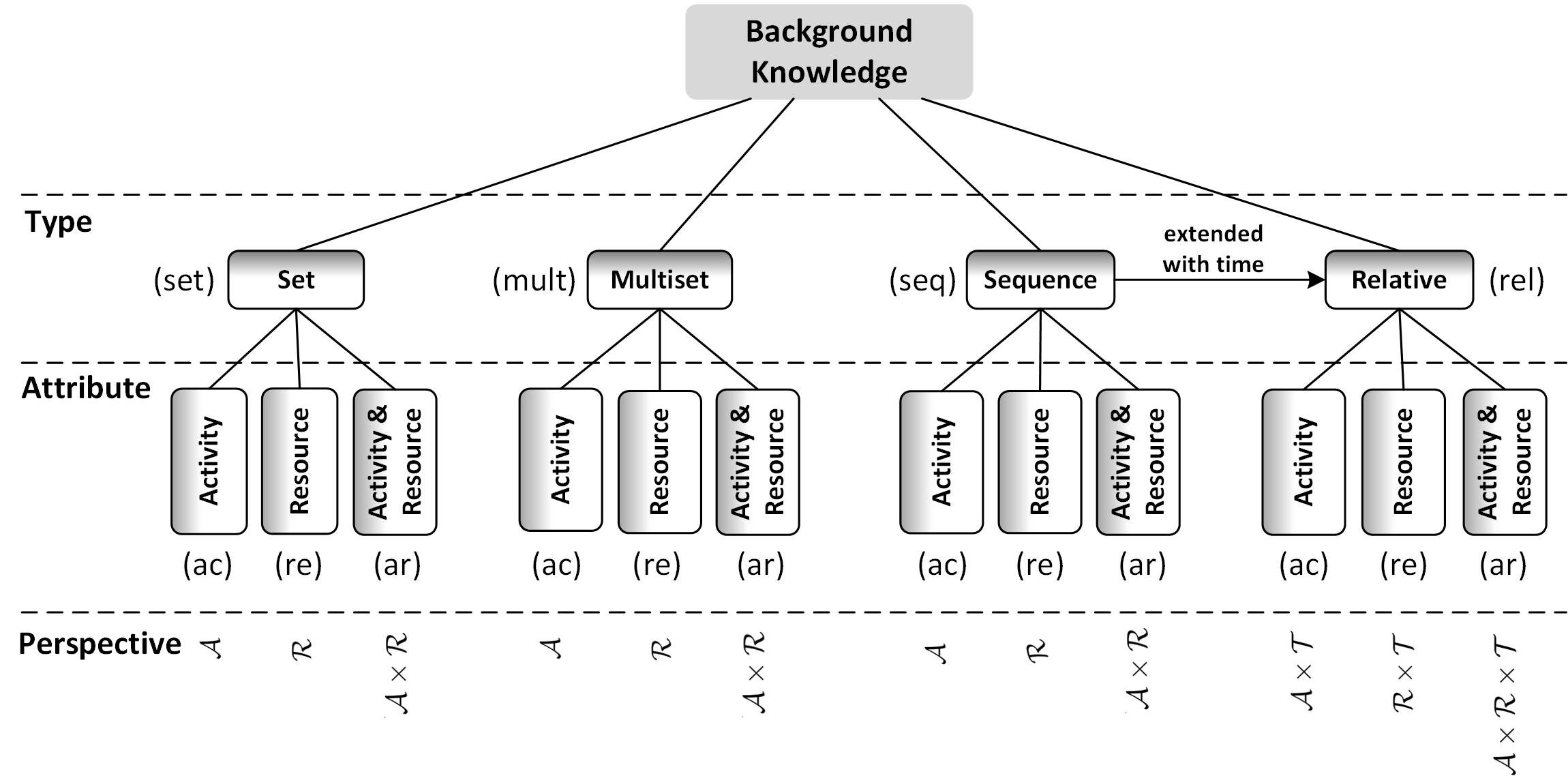}
		\caption{Categorizing background knowledge based on the type and event attributes as well as the corresponding perspectives, e.g., if $type=rel$ and $att=ar$, the corresponding perspective is $ps=\activityUniverse \times \resourceUniverse \times \timeUniverse$.}\label{fig:bk}
	\end{figure}

	Using relative timestamps does not eliminate time-based attacks, since the time differences are real and can be exploited by an adversary. \textit{Relative time difference} type of background knowledge is an extension for the \textit{sequence} type, where the assumption is that an adversary knows a subsequence of some event attributes as well as the relative time differences between the elements. \autoref{fig:bk} shows the classification of background knowledge based on the types and event attributes.
	In the following, we provide formal definitions for different categories of background knowledge based on the main event attributes, i.e., \textit{activity}, \textit{resource}, and \textit{timestamp}.
	Moreover, one can see that there is a relation between \textit{type}, \textit{attribute}, and \textit{perspective}, i.e., a combination of type and attribute can be mapped to a perspective. For example, if $type=rel$ and $att=ar$, the corresponding perspective is $ps=\activityUniverse \times \resourceUniverse \times \timeUniverse$, or if $type \in \{ set,mult,seq \}$ and $att=re$, the corresponding perspective is $ps=\resourceUniverse$.
	
	\begin{definition}[Background Knowledge Based on Activities]
		\label{def:bk_act}
		Let $EL$ be an event log, and $A_{EL}$ be the set of activities in the event log. $bk_{set,ac}(EL)=2^{A_{EL}}$, $bk_{mult,ac}(EL)=\multiset(A_{EL})$, and $bk_{seq,ac}(EL)=A_{EL}^*$ are the sets of candidates of background knowledge based on the activity attribute of the events for the \textit{set},\textit{multiset}, and \textit{sequence} types of background knowledge. 
		For example, $\{ a,b,c \} \in bk_{set,ac}(EL)$, $[a^2,b] \in bk_{mult,ac}(EL)$, and $\langle a,b,c \rangle \in bk_{seq,ac}(EL)$. 
	\end{definition}

	\begin{definition}[Background Knowledge Based on Resources]
		\label{def:bk_res}
		Let $EL$ be an event log, and $R_{EL}$ be the set of activities in the event log. $bk_{set,re}(EL)=2^{R_{EL}}$, $bk_{mult,re}(EL)=\multiset(R_{EL})$, and $bk_{seq,re}(EL)=R_{EL}^*$ are the sets of candidates of background knowledge based on the resource attribute of the events for the different types of background knowledge. 
	\end{definition}

	\begin{definition}[Background Knowledge Based on Activities\&Resources]
		\label{def:bk_act_res}
		Let $EL$ be an event log, $A_{EL}$ be the set of activities in the event log, and $R_{EL}$ be the set of resources in the event log. $bk_{set,ar}(EL)=2^{A_{EL} \times R_{EL}}$, $bk_{mult,ar}(EL)=\multiset(A_{EL} \times R_{EL})$, and $bk_{seq,ar}(EL)=(A_{EL} \times R_{EL})^*$ are the sets of candidates of background knowledge based on the activity and resource attribute of the events for the various types of background knowledge. 
	\end{definition}

	\begin{definition}[Background Knowledge Based on Time Differences Between Relative Timestamps]
		\label{def:bk_time}
		Let $EL$ be an event log, $A_{EL}$ be the set of activities in the event log, $R_{EL}$ be the set of resources in the event log, and $\timeUniverse$ be the universe of (relative) timestamps. $bk_{rel,ac}(EL)= (A_{EL} \times \timeUniverse)^*$, $bk_{rel,re}(EL)=(R_{EL} \times \timeUniverse)^*$, and $bk_{rel,ar}(EL)=(A_{EL} \times R_{EL} \times \timeUniverse)^*$ are the sets of candidates of background knowledge based on the relative time differences. 
	\end{definition}
	
	Note that in \autoref{def:bk_time}, other attributes are also present. However, our focus is on time differences between relative timestamps. Therefore, we refer to this category of background knowledge as time-based.

	\section{Attack Models}\label{sec:attackmodel}
\begin{figure}[bt]
		\centering
		\includegraphics[width=0.95\textwidth]{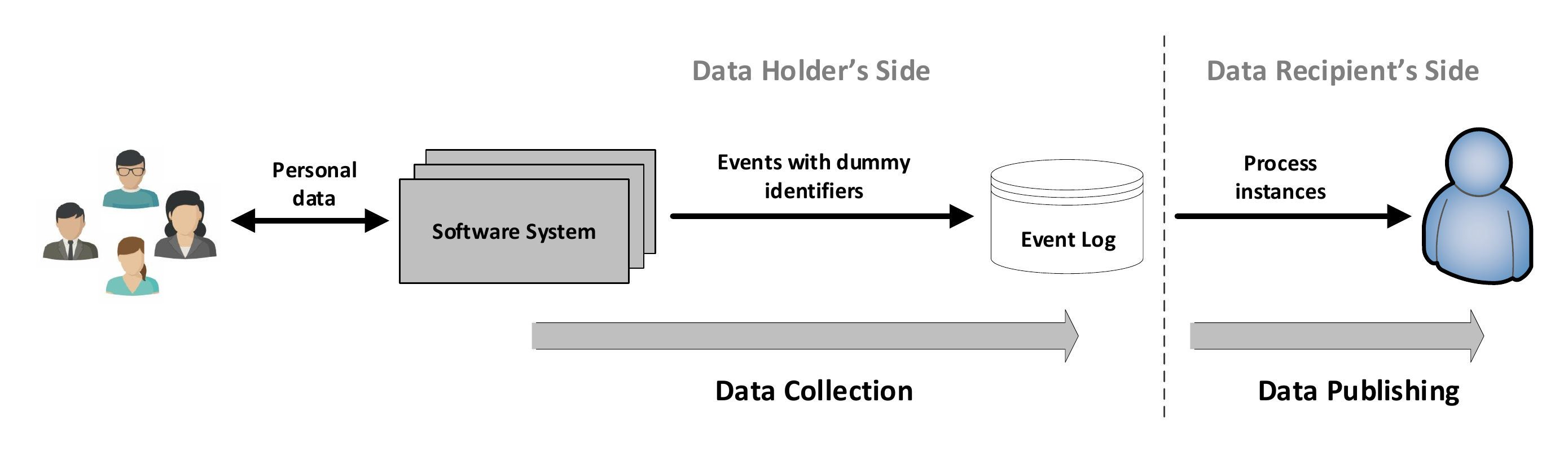}
		\caption{Data collection and data publishing scenario.}\label{fig:scenario}
	\end{figure}

	\autoref{fig:scenario} shows our simple scenario of data collection and data publishing. With respect to the types of data holder's models, introduced in \cite{Gehrke06}, we consider a \textit{trusted model}. In the trusted data holder models, the \textit{data holder} is trustworthy, and on the data holder's side, only simple anonymization techniques need to be applied, e.g., suppressing real identifiers. However, the \textit{data recipient}, i.e., a process miner, is not trustworthy and may attempt to identify sensitive information about record owners, i.e., cases.
	Given a process instance $p=(c,\sigma,s) \in \processUniverse$, both $\sigma$ and $s$ are considered as sensitive person-specific information, and part of the trace $\sigma$ can be exploited as the \textit{quasi-identifier} to re-identify the owner of the process instance, i.e., $c$, and/or to learn the sensitive information which belongs to the data owner, i.e., $\sigma$ and/or $s$.
	
	In the following, we provide formal definitions and examples for the attack scenarios based on the main event attributes, i.e., \textit{activity}, \textit{resource}, and \textit{timestamp}. Note that the examples are based on the event log shown in Table$~\ref{simple_evenlog}$.

	\subsection{Activity-based Attacks}
	In the activity-based scenarios, we assume that the adversary's knowledge is about the activities performed for a victim case. In the following, we provide formal models based on the introduced types of background knowledge. 
	
	\begin{itemize}
		\item \textbf{Based on a set of activities (A1):} In this scenario, we assume that the adversary knows a subset of activities performed for a case, and this information can lead to the \textit{case linkage} and/or \textit{attribute linkage} attacks. 
		Given $EL$ as an event log, we formalize this scenario by a function $match_{set,ac}^{EL}: 2^{A_{EL}} \rightarrow{2^{EL}}$. For $A \in bk_{set,ac}(EL)$, $match_{set,ac}^{EL}(A) = \{(c,\sigma,s) \in EL \mid A \subseteq \{a \in \pi_{\activityUniverse}(\sigma)\}  \}$. 
		For example, if the adversary knows that $\{VI,IN\}$ is a subset of activities performed for a case, the only matching case is case $4$. Therefore, both the sequence of events and the sensitive attribute are disclosed. 
		
		\item \textbf{Based on a multiset of activities (A2):} In this scenario, we assume that the adversary knows a sub-multiset of activities performed for a case, and this information can result in the linkage attacks. 
		Given $EL$ as an event log, we formalize this scenario as follows. $match_{mult,ac}^{EL}: \multiset(A_{EL}) \rightarrow{2^{EL}}$. For $B \in bk_{mult,ac}(EL)$, $match_{mult,ac}^{EL}(B) = \{(c,\sigma,s) \in EL \mid B \subseteq [a \in \pi_{\activityUniverse}(\sigma)] \}$.
		For example, if the adversary knows that $[HO^1, BT^2]$ is a multiset of activities performed for a case, the only matching case is case $2$. Consequently, the complete sequence of events and the disease are disclosed. 
		
		\item \textbf{Based on a sequence of activities (A3):} In this scenario, we assume that the adversary knows a subsequence of activities performed for a case, and this information can lead to the linkage attacks. 
		Given $EL$ as an event log, we formalize this scenario by a function $match_{seq,ac}^{EL}: A_{EL}^* \rightarrow{2^{EL}}$. For $\sigma \in bk_{seq,ac}(EL)$, $match_{seq,ac}^{EL}(\sigma) = \{(c,\sigma',s) \in EL \mid \sigma \sqsubseteq \pi_{\activityUniverse}(\sigma') \}$.
		For example, if the adversary knows that $\langle RE,VI,HO \rangle$ is a subsequence of activities performed for a case, case 5 is the only matching case.
		
	\end{itemize}

	\subsection{Resource-based Attacks}
	In the resource-based scenarios, we assume that the adversary's knowledge is about the resources who perform activities for a victim case. In the following, we provide formal models based on the main types of background knowledge.
	
	\begin{itemize}
		\item \textbf{Based on a set of resources (R1):} In this scenario, we assume that the adversary knows a subset of resources involved in performing activities for a victim case, and this information can lead to the \textit{case linkage} and/or \textit{attribute linkage} attacks.
		Given $EL$ as an event log, we formalize this scenario as follows. $match_{set,re}^{EL}: 2^{R_{EL}} \rightarrow{2^{EL}}$. For $R \in bk_{set,re}(EL)$, $match_{set,re}^{EL}(R) = \{(c,\sigma,s) \in EL \mid R \subseteq \{r \in \pi_{\resourceUniverse}(\sigma)\}  \}$. 
		For example, if the adversary knows that $\{E1,D2\}$ is a subset of resources involved in handling a victim case, case 5 is the only matching case. Therefore, both the sequence of events and the sensitive attribute are disclosed. 
		
		\item \textbf{Based on a multiset of resources (R2):} In this scenario, we assume that the adversary knows a sub-multiset of resources involved in performing activities for a victim case, and this information can lead to the linkage attacks. 
		Given $EL$ as an event log, we formalize this scenario as follows. $match_{mult,re}^{EL}: \multiset(R_{EL}) \rightarrow{2^{EL}}$. For $S \in bk_{mult,re}(EL)$, $match_{mult,re}^{EL}(S) = \{(c,\sigma,s) \in EL \mid S \subseteq [r \in \pi_{\resourceUniverse}(\sigma)] \}$.
		For example, if the adversary knows that $[N1^2,E3]$ is a multiset of resources performed activities for a victim case, the only matching case is case 2. 
		
		\item \textbf{Based on a sequence of resources (R3):} In this scenario, we assume that the adversary knows a subsequence of resources who performed activities for a victim case, and this information can result in the linkage attacks. 
		Given $EL$ as an event log, we formalize this scenario by a function $match_{seq,re}^{EL}: R_{EL}^* \rightarrow{2^{EL}}$. For $\sigma \in bk_{seq,re}(EL)$, $match_{seq,re}^{EL}(\sigma) = \{(c,\sigma',s) \in EL \mid \sigma \sqsubseteq \pi_{\resourceUniverse}(\sigma') \}$.
		For example, if the adversary knows that $\langle E4,D2 \rangle$ is a subsequence of resources who performed activities for a victim case, the only matching case is case 4.
		
	\end{itemize}

	\subsection{Activity\&Resource-based Attacks}
	In the activity\&resource-based scenarios, we assume that the adversary's knowledge is about activities and the corresponding resources who perform activities for a victim case. In the following, we provide formal models based on the main types of background knowledge.
	
	\begin{itemize}
		\item \textbf{Based on a set of (activity,resource) pairs (AR1):} In this scenario, we assume that the adversary knows a subset of (activity,resource) pairs included in the trace attribute of a victim case, and this information can result in the \textit{case linkage} and/or \textit{attribute linkage} attacks. 
		Given $EL$ as an event log, we formalize this scenario as follows. $match_{set,ar}^{EL}: 2^{A_{EL} \times R_{EL}} \rightarrow{2^{EL}}$. For $AR \in bk_{set,ar}(EL)$, $match_{set,ar}^{EL}(AR) = \{(c,\sigma,s) \in EL \mid AR \subseteq \{(a,r) \in \pi_{\activityUniverse \times \resourceUniverse}(\sigma)\}  \}$. 
		For example, if the adversary knows that $\{(HO,E6)\}$ is a subset of (activity,resource) pairs contained in the trace attribute of a victim case, case 5 is the only matching case, which result is the whole sequence and sensitive attribute disclosure.
		
		\item \textbf{Based on a multiset of (activity,resource) pairs (AR2):} In this scenario, we assume that the adversary knows a sub-multiset of (activity,resource) pairs included in the trace attribute of a victim case.
		Given $EL$ as an event log, the scenario can be formalized as follows. $match_{mult,ar}^{EL}: \multiset(A_{EL} \times R_{EL}) \rightarrow{2^{EL}}$. For $BS \in bk_{mult,ar}(EL)$, $match_{mult,ar}^{EL}(BS) = \{(c,\sigma,s) \in EL \mid BS \subseteq [(a,r) \in \pi_{\activityUniverse \times \resourceUniverse}(\sigma)] \}$.
		For example, if the adversary knows that $[(BT,N1)^2]$ is a multiset of (activity,resource) pairs included in the trace attribute of a victim case, the only matching case is case 2.
		
		\item \textbf{Based on a sequence of (activity,resource) pairs (AR3):} In this scenario, we assume that the adversary knows a subsequence of (activity,resource) pairs included in the trace attribute of a victim case, and this information can lead to the linkage attacks. 
		Given $EL$ as an event log, we formalize this scenario by a function $match_{seq,ar}^{EL}: (A_{EL} \times E_{EL})^* \rightarrow{2^{EL}}$. For $\sigma \in bk_{seq,ar}(EL)$, $match_{seq,ar}^{EL}(\sigma) = \{(c,\sigma',s) \in EL \mid \sigma \sqsubseteq \pi_{\activityUniverse \times \resourceUniverse}(\sigma') \}$.
		For example, if the adversary knows that $\langle (RE,E4),(VI,D2) \rangle$ is a (activity,resource) pairs included in the trace attribute of a victim case, case 4 is the only matching case.

	\end{itemize}
	
	\subsection{Time-based Attacks}
	As we discussed in \autoref{subsec:bk}, after making the timestamps relative, the time differences are still real and can be exploited by an adversary. In the following, we extend the attacks of the type \textit{sequence}, i.e., A3, R3, AR3, with the time-related information.

	\begin{itemize}
		\item \textbf{Based on relative time differences between activities (AT):} In this scenario, we assume that the adversary knows a subsequence of activities and also the time difference between the activities. Given $EL$ as an event log, the scenario is formalized as follows.
		$match_{rel,ac}^{EL}: (A_{EL} \times \timeUniverse)^* \rightarrow{2^{EL}}$. For $\sigma \in bk_{rel,ac}(EL)$, $match_{rel,ac}^{EL}(\sigma) = \{(c,\sigma',s) \in EL \mid \sigma \sqsubseteq relative(\pi_{\activityUniverse \times \timeUniverse}(\sigma')) \}$.
		For example, if an adversary's knowledge is $\langle HO,VI \rangle$, both case 2 and case 3 get matched. However, if the adversary further knows that for a victim case, \textit{visit} performed in the morning of the next day, the only matching case is case 2.
		
		\item \textbf{Based on relative time differences between resources who performed activities (RT):} According to this scenario, the adversary knows a subsequence of resources and the time difference between the resources involved in handling a case.
		Given $EL$ as an event log, we formalize this scenario by a function $match_{rel,re}^{EL}: (R_{EL} \times \timeUniverse)^* \rightarrow{2^{EL}}$. For $\sigma \in bk_{rel,re}(EL)$, $match_{rel,re}^{EL}(\sigma) = \{(c,\sigma',s) \in EL \mid \sigma \sqsubseteq relative(\pi_{\resourceUniverse \times \timeUniverse}(\sigma')) \}$.
		For example, if an adversary's knowledge is $\langle E1,E3 \rangle$, both case 2 and case 3 get matched. However, if the adversary further knows that for the victim case, \textit{employee} 3 performed \textit{hospitalization} more than one hour after \textit{registration}, case 3 is the only matching case.
		
		\item \textbf{Based on relative time differences between (activity,resource) pairs (ART):} In this scenario, the assumption is that the adversary knows a subsequence of (activity,resource) pairs and the time difference between these pairs.
		Given $EL$ as an event log, we formalize this scenario as follows. $match_{rel,ar}^{EL}: (A_{EL} \times R_{EL})^* \rightarrow{2^{EL}}$. For $\sigma \in bk_{rel,ar}(EL)$, $match_{rel,ar}^{EL}(\sigma) = \{(c,\sigma',s) \in EL \mid \sigma \sqsubseteq relative(\sigma') \}$.
		For example, case 1 and case 6 have the same sequence of (activity,resource) pairs. However, if the adversary knows that for a victim case, it took almost four hours to get released by \textit{employee} 6 after visiting by a doctor, the corresponding possible cases narrow down to only one case, which is case 6.
		
	\end{itemize}

	\section{Privacy Preservation Techniques}\label{sec:PrivacyModel}
	Traditional $k$-anonymity and its extended privacy preservation techniques assume that an adversary could use all of the quasi-identifier attributes as background knowledge to launch linkage attacks. According to the types of background knowledge introduced in Section~\ref{sec:prelimineries}, this assumption means that the background knowledge of an adversary is $bk_{rel,ar}$ which covers all the information contained in a trace.
	In the following, we show the results of applying two baseline methods with respect to the aforementioned assumption.

	\subsection{Baseline Methods}
	In this subsection, we introduce two baseline methods to apply $k$-anonymity on event logs: \textit{Baseline}-1 and \textit{Baseline}-2. \textit{Baseline}-1 is a na\"ive $k$-anonymity approach where we remove all the trace variants occurring less than $k$ times. \textit{Baseline}-2 maps each violating trace variant, i.e., the variant that does not fulfill the desired $k$-anonymity requirement, to the most similar non-violating subtrace by removing events. In \textit{Baseline}-2, if there exists no non-violating subtrace, the whole trace variant is removed. 
	  
	\begin{table}[t]
		\centering
		\scriptsize
		\caption{A simple event log where time difference between relative timestamps are represented by integer values.}
		\label{tbl:example1}
		\begin{tabular}{|l|l|l|}
			\hline
			Case Id & Trace                                                     & Disease   \\ \hline
			1       & $\langle(RE,E4,1),(HO,E3,4),(VI,D1,5),(BT,N1,7),(VI,D1,8)\rangle$ & Cancer    \\ \hline
			2       & $\langle(BT,N1,7),(VI,D1,8),(RL,E2,9)\rangle$              & Infection \\ \hline
			3       & $\langle(HO,E3,4),(VI,D1,5),(BT,N1,7),(RL,E2,9)\rangle$       & Corona \\ \hline
			4       & $\langle(RE,E4,1),(VI,D1,6),(VI,D1,8),(RL,E2,9)\rangle$        & Infection \\ \hline
			5       & $\langle(HO,4),(VI,D1,8),(RL,E2,9)\rangle$              & Corona \\ \hline
			6       & $\langle(VI,D1,6),(BT,N1,7),(RL,E2,9)\rangle$              & Flu       \\ \hline
			7       & $\langle(RE,E4,1),(BT,N1,7),(VI,D1,8),(RL,E2,9)\rangle$       & Flu       \\ \hline
			8       & $\langle(RE,E4,1),(VI,D1,6),(BT,N1,7),(VI,D1,8)\rangle$        & Cancer    \\ \hline
		\end{tabular}
	\end{table}
	
	Suppose that \autoref{tbl:example1} is part of an event log recorded by an information system in a hospital that needs to be published after applying $k$-anonymity. Note that for the sake of simplicity, the time differences between relative timestamps are represented by integers. Since all the traces in this event log are unique if we apply $k$-anonymity with any value greater than 1, using \textit{Baseline}-1, all the traces are removed. If we apply \textit{Baseline}-2 where $k=2$ then the result is the event log shown in \autoref{tbl:k2}. One can see that for such a weak privacy requirement 12 events are removed. Now, if we use $k=4$, \autoref{tbl:k4} is the result where 18 events are removed which is more than half of the events.

	\begin{table}[b]
		\parbox{.49\linewidth}{
			\centering
			\tiny
			\caption{The event log after applying 2-anonymity to \autoref{tbl:example1} using \textit{Baseline}-2.}\label{tbl:k2}
			\begin{tabular}{|l|l|l|}
			\hline
			Case Id & Trace                                                     & Disease   \\ \hline
			1       & $\langle(BT,N1,7),(VI,D1,8)\rangle$ & Cancer    \\ \hline
			2       & $\langle(BT,N1,7),(VI,D1,8),(RL,E2,9)\rangle$              & Infection \\ \hline
			3       & $\langle(BT,N1,7),(RL,E2,9)\rangle$       & Corona \\ \hline
			4       & $\langle(VI,D1,8),(RL,E2,9)\rangle$        & Infection \\ \hline
			5       & $\langle(VI,D1,8),(RL,E2,9)\rangle$              & Corona \\ \hline
			6       & $\langle(BT,N1,7),(RL,E2,9)\rangle$              & Flu       \\ \hline
			7       & $\langle(BT,N1,7),(VI,D1,8),(RL,E2,9)\rangle$       & Flu       \\ \hline
			8       & $\langle(BT,N1,7),(VI,D1,8)\rangle$        & Cancer    \\ \hline
		\end{tabular}
			
		}
		\hfill
		\parbox{.49\linewidth}{
			\centering
			\tiny
			\caption{The event log after applying 4-anonymity to \autoref{tbl:example1} using \textit{Baseline}-2.}\label{tbl:k4}
			\begin{tabular}{|l|l|l|}
				\hline
				Case Id & Trace                                                     & Disease   \\ \hline
				1       & $\langle(BT,N1,7),(VI,D1,8)\rangle$ & Cancer    \\ \hline
				2       & $\langle(BT,N1,7),(VI,D1,8)\rangle$              & Infection \\ \hline
				3       & $\langle(RL,E2,9)\rangle$       & Corona \\ \hline
				4       & $\langle(RL,E2,9)\rangle$        & Infection \\ \hline
				5       & $\langle(RL,E2,9)\rangle$              & Corona \\ \hline
				6       & $\langle(RL,E2,9)\rangle$              & Flu       \\ \hline
				7       & $\langle(BT,N1,7),(VI,D1,8)\rangle$       & Flu       \\ \hline
				8       & $\langle(BT,N1,7),(VI,D1,8)\rangle$        & Cancer    \\ \hline
			\end{tabular}
		}
	\end{table}

	In \cite{pretsaICPM2019}, the $PRETSA$ method is introduced as a group-based privacy preservation technique for process mining where the authors apply $k$-anonymity and $t$-closeness on event data for privacy-aware process discovery. However, $PRETSA$ focuses on the \textit{resource perspective} of privacy while we focus on the \textit{case perspective}. The $PRETSA$ method assumes a prefix of activity sequences as the background knowledge, and each violating trace is mapped to the most similar non-violating trace. In \cite{rafieitlkc}, $PRETSA_{case}$ is introduced as a variant of $PRETSA$ method where only the $k$-anonymity part is considered, and the focus is on the privacy of \textit{cases} rather than \textit{resources}. Therefore, $PRETSA_{case}$ is a specific type of \textit{Baseline}-2 where the background knowledge is a specific type of $bk_{seq,ac}$, i.e., a prefix of activity sequences rather than any subsequence.

	\subsection{$\tlkc$-Privacy (Extended)}

	As discussed in \cite{rafieitlkc}, it is almost impossible for an adversary to acquire all the information of a target victim, and it requires non-trivial effort to gather each piece of background knowledge. The $\tlkc$-privacy exploits this limitation and assumes that the adversary's background knowledge is bounded by at most $L$ values of the quasi-identifier, i.e., the size or power of background knowledge. Based on the types of background knowledge illustrated in \autoref{fig:bk}, the $\tlkc$-privacy considers all the types, i.e., \textit{set}, \textit{multiset}, \textit{sequence}, and \textit{relative}. However, it focuses on the \textit{activity} attribute (ac) and \textit{timestamps} which are included in the \textit{relative} type. In this paper, the technique is extended with the \textit{resource} attribute, i.e., merely \textit{resource} (re) and \textit{activity} along with \textit{resource} (ar) are also considered. In the following, we bound the power of the different types of background knowledge (\autoref{def:bk_act}-\ref{def:bk_time}) with $L$ as the maximal size of candidates.
	
	\begin{definition}[Bounded Background Knowledge]
		\label{def:bounded_bk}
		Let $EL$ be an event log, $type \in \{ set,mult,seq,rel \}$ be the type of background knowledge, $att \in \{ ac,re,ar \}$ be the event attribute of background knowledge, and L be the size of background knowledge. $bk_{type,att}^L(EL)=\{ cand \in bk_{type,att}(EL) \mid |cand| \le L \}$ are the candidates of the background knowledge whose sizes are bounded by $L$.  
	\end{definition}
	
	In the $\tlkc$-privacy, $T \in \{ seconds, minutes, hours, days \}$ refers to the accuracy of timestamps, e.g., $T=minutes$ shows that the accuracy of timestamps is limited at \textit{minutes} level, $L$ refers to the power of background knowledge, $K$ refers to the $k$ in the $k$-anonymity definition, and $C$ refers to the bound of confidence regarding the sensitive attribute values in a matching set.
	We denote $EL(T)$ as the event log with the accuracy of timestamps at the level $T$. The general idea of $TLKC$-privacy is to ensure that the background knowledge of size $L$ in $EL(T)$ is shared by at least $K$ cases, and the confidence of inferring the sensitive value in $S$ is not greater than $C$.

	\begin{definition}[$TLKC$-Privacy]
		\label{def:tlkc-privacy}
		Let $EL \subseteq \processUniverse$ be an event log, $L$ be the maximal size of background knowledge, $T \in \{ seconds, minutes, hours, days \}$ be the accuracy of timestamps, $type \in \{ set, mult, seq, rel \}$, and $att \in \{ ac,re,ar \}$. $EL(T)$ satisfies $TLKC$-privacy if and only if for any $cand \in bk_{type,att}^L(EL(T))$ such that $match_{type,att}^{EL(T)}(cand) \ne \emptyset$:
		\begin{itemize}
			\item $|match_{type,att}^{EL(T)}(cand)| \ge K$, where $K \in \mathbb{N}_{>0}$, and
			\item $Pr(s|cand) = \frac{|\{ p \in match_{type,att}^{EL(T)}(cand) \mid \pi_{\sensitiveUniverse}(p)=s \}|}{|match_{type,att}^{EL(T)}(cand)|} \le C$ for any $s \in S$, where $0 < C \le 1$ is a real number as the confidence threshold, and $\pi_{\sensitiveUniverse}(p)$ is the projection of the process instance on the sensitive attribute value.
		\end{itemize}	
		
	\end{definition}

	The $\tlkc$-privacy provides a major relaxation from traditional $k$-anonymity based on a reasonable assumption that the adversary has restricted knowledge. It generalizes several privacy preservation techniques including $k$-anonymity, confidence bounding, $(\alpha,k)$-anonymity, and $l$-diversity. It also provides interpretable parameters. Note that the type and attribute of background knowledge implicitly show the perspective (Figure~\ref{fig:bk}). 
	
	\subsubsection{Privacy Measure}
	In the subsection, we define \textit{(minimal) violating traces} w.r.t. the privacy requirements of the $\tlkc$-privacy.
	
	\begin{definition}[Violating Trace]
		\label{def:violatingSequence}
		Let $EL \subseteq \processUniverse$ be an event log, $L$ be the maximal size of background knowledge, $T \in \{ seconds, minutes, hours, days \}$ be the accuracy of timestamps, $type \in \{ set, mult, seq, rel \}$, $att \in \{ ac,re,ar \}$, $ps \in \perspectiveUniverse$ be the corresponding perspective w.r.t. the given $type$ and $att$, and $\sigma \sqsubseteq \pi_{ps}(\sigma')$ such that $(c,\sigma',s) \in EL(T)$. $\sigma$ is a violating (sub)trace with respect to the $TLKC$-privacy requirements if there exists a $cand \in bk_{type,att}^L(EL(T))$:
		\begin{itemize}
			\item $cand \sqsubseteq \sigma \vee cand \subseteq \{e \in \sigma\} \vee cand \subseteq [e \in \sigma]$, and
			\item $|match_{type,att}^{EL(T)}(cand)| < K$ or $Pr(s|cand) > C$ for some $s \in S$.
		\end{itemize}	
	\end{definition} 

	An event log satisfies $\tlkc$-privacy, if all violating traces w.r.t. the given privacy requirement are removed. A na\"{\i}ve approach is to determine all violating traces and remove them. However, this approach is inefficient due to the numerous number of violating traces, even for a weak privacy requirement.  
	Moreover, as demonstrated in \cite{rafieitlkc}, $\tlkc$-privacy is not monotonic w.r.t. $L$. In fact, the anonymity threshold $K$ is monotonic w.r.t. $L$, i.e., if $L' \le L$ and $C=100\%$, an event log $EL$ which satisfies $TLKC$-privacy must satisfy $TL'KC$-privacy. However, confidence threshold $C$ is not monotonic w.r.t. $L$, i.e., if $\sigma$ is non-violating trace, its subtrace may or may not be non-violating. Therefore, we have to make sure that the conditions should hold for any $L' \le L$. To this end, in the following, we define the extended version of \textit{minimal violating traces} w.r.t. the different perspectives.   
	
	\begin{sloppy}
		\begin{definition}[Minimal Violating Trace]
			\label{def:mvt}
			Let ${EL \subseteq \processUniverse}$ be an event log, $L$ be the maximal size of background knowledge, $T \in \{ seconds, minutes, hours, days \}$ be the accuracy of timestamps, $type \in \{ set, mult, seq, rel \}$, $att \in \{ ac,re,ar \}$, $ps \in \perspectiveUniverse$ be the corresponding perspective w.r.t. the given $type$ and $att$, and $\sigma \sqsubseteq \pi_{ps}(\sigma')$ such that $(c,\sigma',s) \in EL(T)$. 
			$\sigma$ is a minimal violating trace if $\sigma$ is a violating trace $(\autoref{def:violatingSequence})$ in the $EL$, and every proper subtrace of $\sigma$ is not violating.
			We denote $MVT^{EL}_{ps}$ as the set of minimal violating traces in the event log $EL$ w.r.t. the perspective $ps$.  
		\end{definition}
	\end{sloppy}  
	
	Every violating trace in an event log is either a minimal violating trace or it contains a minimal violating trace. Therefore, if an event log contains no minimal violating trace, then it contains no violating trace. Note that the set of minimal violating traces in an event log is much smaller than the set of violating traces in the event log which results in better efficiency for removing violating traces. 
	
	\subsubsection{Utility Measure}\label{subsec:utilitymeasure}
	In the $\tlkc$-privacy, the \textit{maximal frequent traces} are defined as a measure for considering data utility, where traces contain \textit{activity} and \textit{timestamp} attributes. Since we extend the $\tlkc$-privacy preservation technique to cover all the main perspectives of process mining, the utility measure also needs to be extended. In the following, we provide an extended version of the utility measure considering the perspectives. 
	

	\begin{definition}[Maximal Frequent Trace]
		\label{def:mft}
		Let $EL$ be an event log, and $ps \in \perspectiveUniverse$ be a perspective. For a given minimum support threshold $\Theta$, a non-empty trace $\sigma \sqsubseteq \pi_{ps}(\sigma')$ such that $(c,\sigma',s) \in EL$ is \textit{maximal frequent} in the $EL$ if $\sigma$ is frequent, i.e., the frequency of $\sigma$ is greater than or equal to $\Theta$, and no supertrace of $\sigma$ is frequent in the $EL$. We denote $MFT^{EL}_{ps}$ as the set of maximal frequent traces in the event log $EL$ w.r.t. the perspective $ps$.
	\end{definition}
	
	
	The goal of data utility is to preserve as many MFT as possible w.r.t. the given perspective. For example, in the \textit{control-flow} perspective, i.e., $ps=\activityUniverse$, the goal in to preserve the maximal frequent traces w.r.t. the activities. Note that in an event log, the set of maximal frequent traces is much smaller than the set of frequent traces. Moreover, any subtrace of a maximal frequent trace is also a frequent trace, and once all the MFTs are discovered, the support counts of any frequent subtrace can be computed by scanning the data once.

%

	\subsubsection{Balancing Privacy and Utility}\label{subsec:score}
	As discussed in the privacy measure section, to provide the desired privacy requirements, all the minimal violating traces need to be removed. However, this should be done w.r.t. the utility measure. According to \autoref{def:mvt}, every proper subtrace of a minimal violating trace is not violating. Therefore, a minimal violating trace can be removed after removing one event of the trace. This event needs to be chosen w.r.t. both utility and privacy measures. 
	To this end, a greedy function is defined to choose an event to remove from the minimal violating traces such that it maximizes the number of removed minimal violating traces, i.e., privacy gain, yet, at the same time, minimizes the number of removed maximal frequent traces, i.e., utility loss.

	\begin{algorithm}[t]
		\scriptsize
		\SetAlgoLined
		\KwIn{Original event log $EL$}
		\KwIn{$T$, $L$, $K$, $C$, and $\Theta$ (frequency threshold)}
		\KwIn{Background knowledge type and attribute ($bk_{type,att}$), sensitive attributes $\sensitiveUniverse$}
		\KwOut{Anonymized event log $EL'$ which satisfies the desired $TLKC$-privacy requirements}
		generate $MFT^{EL}_{ps}$ and $MVT^{EL}_{ps}$\;
		generate $MFT^{tree}_{ps}$ and $MVT^{tree}_{ps}$ as the prefix trees for $MFT^{EL}_{ps}$ and $MVT^{EL}_{ps}$\;
		\While{there is node (event) in $MVT^{tree}_{ps}$}{
			select an event (node) $e_w$ that has the highest score to suppress based on $socre(e)^{EL}_{ps}$\;
			delete all the MVTs and MFTs containing the event $e_w$ from $MVT^{tree}_{ps}$ and $MFT^{tree}_{ps}$\;
			update $socre(e)^{EL}_{ps}$ for all the remaining events (nodes) in $MVT^{tree}_{ps}$\;
			add $e_w$ to the suppression set $Sup^{EL}$\;
		}
		\ForEach{$e \in Sup^{EL}$}{suppress all instances of $e$ from $EL$\;}
		return suppressed $EL$ as $EL'$\;
		\caption{$\tlkc$-privacy - extended w.r.t. the different perspectives.}
		\label{alg:tlkc}
	\end{algorithm}

	\begin{definition}[Score, Privacy Gain, Utility Loss]
		\label{def:score}
		Let $EL$ be an event log, $ps \in \perspectiveUniverse$ be a perspective, and
		$events_{ps}(EL)=\{ e \in \pi_{ps}(\sigma) \mid (c,\sigma,s) \in EL \}$ be the set of events in the event log w.r.t. the given perspective.
		$score_{ps}^{EL}: \mainEventUniverse \nrightarrow{\mathbb{R}_{\textgreater 0}}$ is a function which retrieves the score of the events in the event log w.r.t. the perspective. 
		For $e \in events_{ps}(EL)$, $score_{ps}^{EL}(e)=\nicefrac{PG^{EL}_{ps}(e)}{UL^{EL}_{ps}(e)+1}$.
		$PG^{EL}_{ps}(e)$ is the number of MVTs containing the event $e$, i.e., $PG^{EL}_{ps}(e)=|\{ x \in MVT_{ps}^{EL} \mid e \in x \}|$ and $UL^{EL}_{ps}(e)$ is the number of MFTs containing the event $e$, i.e., $UL^{EL}_{ps}(e)=|\{ x \in MFT_{ps}^{EL} \mid e \in x \}|$.
	\end{definition}

	Note that in the score (\autoref{def:score}), 1 is added to the denominator to avoid diving by zero (when $e$ does not belong to any MFT). The event $e$ with the highest score is called the $winner$ event, denoted by $e_w$. Algorithm~\ref{alg:tlkc} summarizes all the steps of $TLKC$-privacy. In the following, we show how the algorithm works on the event log \autoref{tbl:example1}.

	\begin{figure}[t]
		\centering
		\subfloat[$MFT^{tree}_{ps}$]{\includegraphics[width=0.99\textwidth]{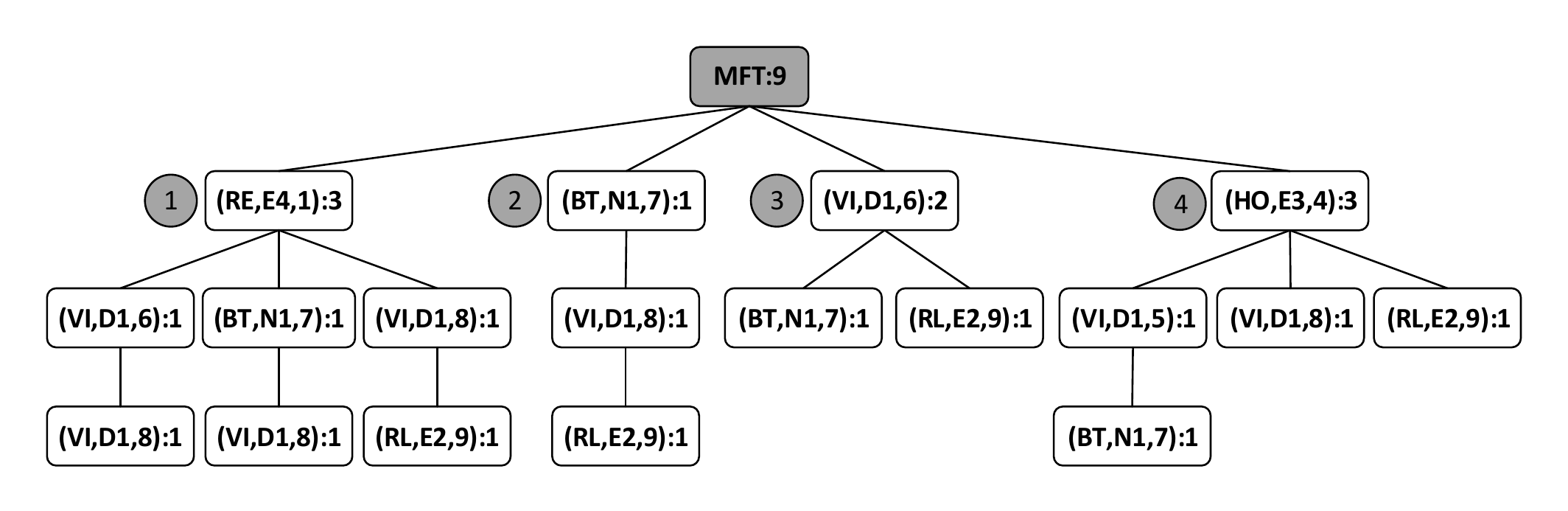}\label{fig:MFT}}
		\hfill
		\subfloat[$MVT^{tree}_{ps}$]{\includegraphics[width=0.60\textwidth]{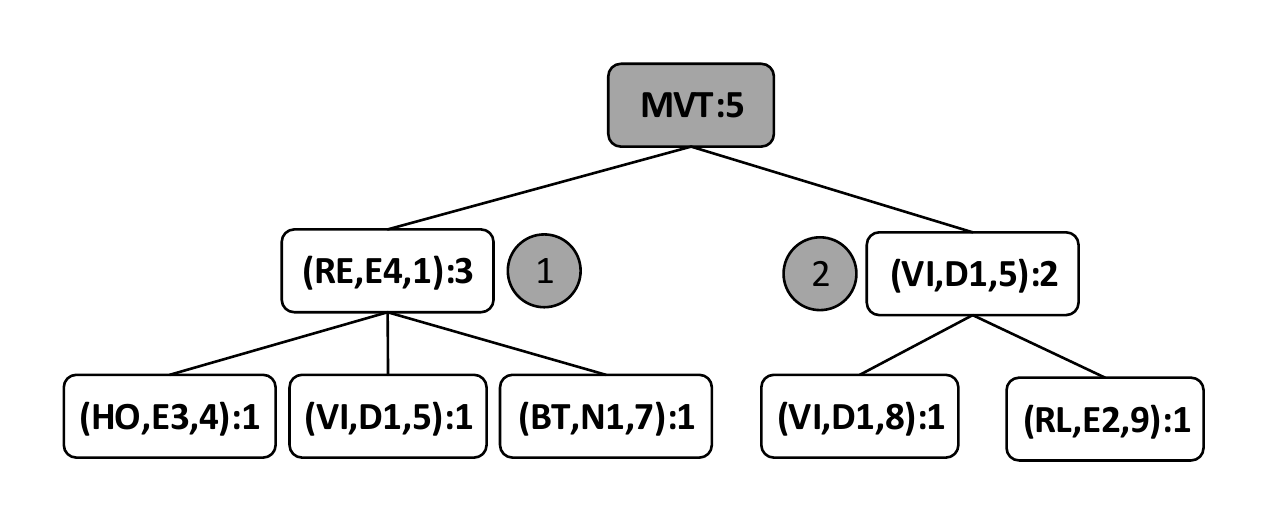}\label{fig:MVT}} \hfill
		\caption{The $MFT^{tree}_{ps}$ and $MVT^{tree}_{ps}$ generated for the event log Table~\ref{tbl:example1} with $T=hours$, $L=2$, $K=2$, $C=50\%$, $\Theta=25\%$, $\sensitiveUniverse=Disease$, and $bk_{rel,ar}^{EL}$.}
		\label{fig:trees}
	\end{figure}

	Suppose that Table~\ref{tbl:example1} shows a simple event log $EL$ where timestamps are represented by integer values as hours. The first line in Algorithm~\ref{alg:tlkc} generates the set of maximal frequent traces ($MFT^{EL}_{ps}$) and the set of minimal violating traces ($MVT^{EL}_{ps}$) from the event log $EL$ with $T=hours$, $L=2$, $K=2$, $C=50\%$, $\Theta=25\%$, \textit{Disease} as the sensitive attribute $\sensitiveUniverse$, and $bk_{rel,ar}^{EL}$ as the background knowledge, i.e., $ps = \activityUniverse \times \resourceUniverse \times \timeUniverse$. 	
	Figure~\ref{fig:trees} shows the $MFT^{tree}_{ps}$ and $MVT^{tree}_{ps}$ generated by line 2 in Algorithm~\ref{alg:tlkc}, where each root-to-leaf path represents one trace, and each node represents an event in a trace with the frequency of occurrence. Table~\ref{tbl:score} shows the initial score of every event (node) in the $MVT^{tree}_{ps}$ ($score^{EL}_{ps}(e)$). Line 4 determines the winner event $e_w$ which is $(VI,D1,5)$. Line 5 deletes all the MVTs and MFTs containing the winner event $e_w$, i.e., subtree 2 and the path $\langle(RE,E4,1),(VI,D1,5)\rangle$ of subtree 1 in the $MVT^{tree}_{ps}$, and the path $\langle(HO,E3,4),(VI,D1,5),(BT,N1,7)\rangle$ of subtree 4 in the $MFT^{tree}_{ps}$ are removed and frequencies get updated. Line 6 updates the scores based on the new frequencies of events. Table~\ref{tbl:score2} shows the remaining events in $MVT^{tree}_{ps}$ with the updated scores. Line 7 adds the winner event to a suppression set $Sup^{EL}$. Lines 4-7 is repeated until there is no node in $MVT^{tree}_{ps}$. According to Table~\ref{tbl:score2} the next winner event is $(RE,E4,1)$, and after deleting all the MVTs and MFTs containing this event, $MVT^{tree}_{ps}$ is empty. Therefore, at the end of the \textit{\textbf{while}} loop, the suppression set $Sup^{EL} = \{(VI,D1,5),(RE,E4,1)\}$. The \textit{\textbf{foreach}} loop suppresses all the instances of the events, i.e., \textit{global suppression}, in the $Sup^{EL}$ from the $EL$, and the last line returns the suppressed $EL$ as the anonymized event log $EL'$ which is shown in Table~\ref{tbl:anonymized}.
	
	Compared to the \autoref{tbl:k2} and \autoref{tbl:k4} which are the results of applying traditional $k$-anonymity using \textit{Baseline}-2, Table~\ref{tbl:anonymized} shows that $\tlkc$-privacy removes less events (only 6), for the stronger privacy requirements.

	\begin{table}[t]
		\centering
		\scriptsize
		\caption{The initial scores for the events in Fig.~\ref{fig:MVT}.}
		\label{tbl:score}
		\begin{tabular}{|l|c|c|c|c|c|c|}
			\hline
			& $(RE,E4,1)$ & $(HO,E3,4)$ & $(VI,D1,5)$ & $(BT,N1,7)$ & $(VI,D1,8)$ & $(RL,E2,9)$ \\ \hline
			$PG^{EL}_{ps}(e)$ & 3        & 1        & 3       & 1        & 1       & 1        \\ \hline
			$UL^{EL}_{ps}(e)$+1 & 4        & 4        & 2       & 5        & 6       & 5        \\ \hline
			$score^{EL}_{ps}(e)$       & 0.75     & 0.25     & 1.50     & 0.20      & 0.16    & 0.20      \\ \hline
		\end{tabular}
	\end{table}

	\begin{table}[t]
		\centering
		\scriptsize
		\caption{The first updated scores.}
		\label{tbl:score2}
			\begin{tabular}{|l|c|c|c|}
			\hline
			& $(RE,E4,1)$ & $(HO,E3,4)$ & $(BT,N1,7)$ \\ \hline
			$PG^{EL}_{ps}(e)$ & 2        & 1        & 1        \\ \hline
			$UL^{EL}_{ps}(e)$+1 & 4        & 3        & 4        \\ \hline
			$score^{EL}_{ps}(e)$       & 0.5      & 0.33     & 0.25     \\ \hline
		\end{tabular}
	\end{table}

%

	\begin{table}[t]
		\centering
		\scriptsize
		\caption{The anonymized event log for Table~\ref{tbl:example1} with $T=hours$, $L=2$, $K=2$, $C=50\%$, $\Theta=25\%$, $\sensitiveUniverse=Disease$, and $bk_{rel,ar}^{EL}$.}
		\label{tbl:anonymized}
		\begin{tabular}{|l|l|l|}
			\hline
			Case Id & Trace                                                     & Disease   \\ \hline
			1       & $\langle(HO,E3,4),(BT,N1,7),(VI,D1,8)\rangle$ & Cancer    \\ \hline
			2       & $\langle(BT,N1,7),(VI,D1,8),(RL,E2,9)\rangle$              & Infection \\ \hline
			3       & $\langle(HO,E3,4),(BT,N1,7),(RL,E2,9)\rangle$       & Corona \\ \hline
			4       & $\langle(VI,D1,6),(VI,D1,8),(RL,E2,9)\rangle$        & Infection \\ \hline
			5       & $\langle(HO,E3,4),(VI,D1,8),(RL,E2,9)\rangle$              & Corona \\ \hline
			6       & $\langle(VI,D1,6),(BT,N1,7),(RL,E2,9)\rangle$              & Flu       \\ \hline
			7       & $\langle(BT,N1,7),(VI,D1,8),(RL,E2,9)\rangle$       & Flu       \\ \hline
			8       & $\langle(VI,D1,6),(BT,N1,7),(VI,D1,8)\rangle$        & Cancer    \\ \hline
		\end{tabular}
	\end{table}

	\subsubsection{New Utility Measure and New Score}
	In this subsection, we first describe the shortcomings of the utility measure and the score introduced in \cite{rafieitlkc} (extended in \autoref{def:mft} and \autoref{def:score}), then we introduce a new utility measure and a new score to overcome the drawbacks. 
	According to \autoref{def:score}, the score is calculated based on the existence of events in the set of minimal violating traces and the set of maximal frequent traces. However, the sizes of these sets, and consequently the included events, highly depends on the corresponding parameters. The set of MVTs is obtained based on $T$, $L$, $K$, $C$, and $bk_{type,att}$, while the set of MFTs is discovered based on $\Theta$ and the given perspective.
	Therefore, some of the events included in the set of minimal violating traces may not be included in the set of maximal frequent traces. Consequently, the score of the corresponding events is merely calculated based on the effect on the \textit{privacy gain}. When two or more events have the same score based on the \textit{privacy gain}, the algorithm assumes an equal effect for the data utility aspect and randomly choose one of the events, which is not a valid assumption. 
	
	Another problem with the current score is that even when there are maximal frequent traces where the event is included, the score does not differentiate the corresponding MFTs based on their frequencies in the event log. For example, suppose that for two events $e_1$ and $e_2$ in the minimal violating traces there are two maximal frequent traces $MFT_1$ and $MFT_2$ such that $e_1$ is only included in $MFT_1$, i.e., $UL(e_1)=1$, and $e_2$ is only included in $MFT_2$, i.e., $UL(e_2)=1$. Hence, both events get the same score for the utility aspect. However, the corresponding MFTs may have completely different frequencies in the event log which leads to a different impact on the utility. Particularly, this issue is highlighted when the frequency threshold ($\Theta$) is rather low. For example, if $\Theta=50\%$, then frequency of $MFT_1$ and $MFT_2$ in the event log could differ up to 50\%. Furthermore, the current score is not normalized, and it is not possible for the user to adjust the effect of each aspect on the score. For example, one may want to consider more effect for the data utility aspect compared to the privacy gain aspect.
	
	To overcome the above-mentioned shortcomings, we define a new utility measure that is able to show the impact of every single event on the data utility. We also define a new score based on the new utility measure which provides normalized scores, and the effect of each aspect is adjustable for users. In the new utility measure (\autoref{def:new_utility}), we consider the relative frequency of the variants, where the given perspective of the event is included, as the basis of the utility.

	\begin{definition}[New Utility Measure]
		\label{def:new_utility}
		Let $EL$ be an event log, $ps \in \perspectiveUniverse$ be a perspective, and $events_{ps}(EL)=\{ e \in \pi_{ps}(\sigma) \mid (c,\sigma,s) \in EL \}$ be the set of events in the event log w.r.t. the given perspective. 
		For $e \in events_{ps}(EL)$, $nUL^{EL}_{ps}(e) = 1 - \sum_{\{\sigma \in \widetilde{EL} \mid e \in \pi_{ps}(\sigma)\}}freq^{EL}(\sigma)$.
	\end{definition}

	\begin{definition}[New Score]
		\label{def:new_score}
		Let $EL$ be an event log, $ps \in \perspectiveUniverse$ be a perspective, $events_{ps}(EL)=\{ e \in \pi_{ps}(\sigma) \mid (c,\sigma,s) \in EL \}$ be the set of events in the event log w.r.t. the given perspective, $\alpha$ be the coefficient of privacy gain $(0 \le \alpha \leq 1)$, $\beta$ be the coefficient of utility loss $(0\le \beta \leq 1)$, and $\alpha + \beta = 1$. $n\mbox{-}score_{ps}^{EL}: \mainEventUniverse \nrightarrow{\mathbb{R}_{\textgreater 0}}$ is a function which retrieves the score of the events in the event log w.r.t. the perspective. 
		For $e \in events_{ps}(EL)$, $n\mbox{-}score_{ps}^{EL}(e)=\alpha \cdot rPG^{EL}_{ps}(e) + \beta \cdot nUL^{EL}_{ps}(e)$, where
		$rPG^{EL}_{ps}(e)$ is the relative value of the privacy gain, i.e., $rPG^{EL}_{ps}(e)=\nicefrac{|\{ x \in MVT_{ps}^{EL} \mid e \in x \}|}{|MVT^{EL}_{ps}|}$.
	\end{definition}

	\begin{algorithm}[t]
		\scriptsize
		\SetAlgoLined
		\KwIn{Original event log $EL$}
		\KwIn{$T$, $L$, $K$, $C$}
		\KwIn{Background knowledge type and attribute ($bk_{type,att}$), sensitive attributes $\sensitiveUniverse$}
		\KwOut{Anonymized event log $EL'$ which satisfies the desired $TLKC$-privacy requirements}
		generate $MVT^{EL}_{ps}$ and $MVT^{tree}_{ps}$\;
		\While{there is node (event) in $MVT^{tree}_{ps}$}{
			select an event (node) $e_w$ that has the highest score to suppress based on $n\mbox{-}socre(e)^{EL}_{ps}$\;
			delete all the MVTs containing the event $e_w$ from $MVT^{tree}_{ps}$\;
			update $n\mbox{-}socre(e)^{EL}_{ps}$ for all the remaining events (nodes) in $MVT^{tree}_{ps}$\;
			add $e_w$ to the suppression set $Sup^{EL}$\;
		}
		\ForEach{$e \in Sup^{EL}$}{suppress all instances of $e$ from $EL$\;}
		return suppressed $EL$ as $EL'$\;
		\caption{$\tlkc$-privacy - extended w.r.t. the different perspectives, new score, and new utility measure.}
		\label{alg:tlkc2}
	\end{algorithm}

	Algorithm~\ref{alg:tlkc2} shows the new algorithm based on the new utility measure and new score, where maximal frequent traces are not used anymore, and the score of events included in the minimal violating traces is calculated based on the new score.
	Note that in both Algorithm~\ref{alg:tlkc} and Algorithm~\ref{alg:tlkc2} the perspective is derived from the background knowledge type and attribute (\autoref{fig:bk}).

	\section{Experiments}\label{sec:exp}
	In this section, we evaluate the extended $\tlkc$-privacy by applying it to real-life event logs. We explore the effect of applying the technique on both \textit{data utility} and \textit{result utility}. The results are also compared with the baseline methods.
	The \textit{result utility} analysis evaluates the similarity of the specific results obtained from the privacy-aware event log with the same type of results obtained from the original event log, while the \textit{data utility} analysis compares the privacy-aware event log with the original event log. As discussed in \cite{rafiei_quantification}, the result utility analysis is highly dependent on the underlying algorithm generating specific results, and the data utility analysis provides a more general evaluation. We perform the evaluation for the three main perspectives including \textit{control-flow}, \textit{organizational}, and \textit{time} perspectives. For the result utility analysis, in each perspective, we focus on a specific type of results. For the control-flow perspective, we focus on \textit{process discovery}, for the organizational perspective, we perform \textit{social network discovery}, and for the time perspective, we perform \textit{bottleneck analysis}. The implementation as a Python program is available on GitHub.\footnote{https://github.com/m4jidRafiei/TLKC-Privacy-Ext}

	\subsection{Experimental Setup}\label{sec:exp_setup}
	
	For the experiments, we employ two human-centered event logs, where the \textit{case identifiers} refer to individuals: Sepsis-Cases, BPIC-2012-APP, and BPIC-2017-APP. Sepsis-Cases \cite{Sepcis_2016_Felix} is a real-life event log containing events of sepsis cases from a hospital. BPIC-2012-APP \cite{BPIC2012} is also a real-life event log about a loan application process taken from a Dutch financial institute. BPIC-2017-APP also pertains to a loan application process of a Dutch financial institute. Table~\ref{tbl:general_statistics} shows the general statistics of these event logs.	
	The Sepsis-Cases event log was included in the experiments because it has some challenging features for privacy preservation techniques, namely, 80\% of traces are unique based on the activity perspective which imposes significant challenges for privacy-preserving process discovery algorithms \cite{rafieitlkc,pretsaICPM2019,MannhardtKBWM19}.
	BPIC-2017-APP has similar properties w.r.t. the resource perspective, i.e., 76\% of traces are unique w.r.t. the resource perspective.
	Note that Sepsis-Cases does not contain resource information and cannot be used for the \textit{organizational} perspective analysis. We employ BPIC-2012-APP and BPIC-2017-APP for the \textit{organizational} perspective.
	Table~\ref{tbl:variant_statistics} shows some statistics about the variants with respect to different perspectives. For example, as mentioned, in Sepsis-Cases, 80\% of traces are unique from the activity perspective, or in BPIC-2017-APP, 76\% of traces are unique from the resource perspective. 
	
		\begin{table}[tb]
		\centering
		\scriptsize
		\caption{The general statistics of the event logs used in the experiments.}\label{tbl:general_statistics}
		\begin{tabular}{|l|l|l|c|c|c|c|c|}
			\hline
			\multicolumn{3}{|c|}{\textbf{Event Log}} & \#cases & \#events & \begin{tabular}[c]{@{}c@{}}\#unique\\activity\end{tabular} & \begin{tabular}[c]{@{}c@{}}\#unique\\resource\end{tabular} & \begin{tabular}[c]{@{}c@{}}\#unique\\(activity,resource)\end{tabular} \\ \hline
			\multicolumn{3}{|l|}{Sepsis-Cases \cite{Sepcis_2016_Felix}}       & 1050     & 15214        & 16    & -                                                              & -                                                         \\ \hline
			\multicolumn{3}{|l|}{BPIC-2012-APP \cite{BPIC2012}}      & 13087    & 60849        & 10   & 61                                                              & 301                                                        \\ \hline
			\multicolumn{3}{|l|}{BPIC-2017-APP \cite{BPIC2017}}      &  31509    &  239595        &  10   &  144                                                              & 927                                                        \\ \hline
		\end{tabular}
	\end{table}

	\begin{table}[b]
		\centering
		\scriptsize
		\caption{Some statistics regarding the variants of the event logs used in the experiments w.r.t. the different perspectives.}\label{tbl:variant_statistics}
		\begin{tabular}{|l|l|l|c|c|c|c|c|}
			\hline
			\multicolumn{3}{|c|}{\textbf{Event Log}} & \begin{tabular}[c]{@{}c@{}}\#variants\\activity perspective\end{tabular} & \begin{tabular}[c]{@{}c@{}}\#variants\\resource perspective\end{tabular} & \begin{tabular}[c]{@{}c@{}}\#variants\\(activity,resource) perspective\end{tabular} \\ \hline
			\multicolumn{3}{|l|}{Sepsis-Cases \cite{Sepcis_2016_Felix}}     & 846    & -                                                              & -                                                         \\ \hline
			\multicolumn{3}{|l|}{BPIC-2012-APP \cite{BPIC2012}}        & 17   & 2974                                                              & 3872                                                        \\ \hline
			\multicolumn{3}{|l|}{BPIC-2017-APP \cite{BPIC2017}}        & 102   &  24230                                                              &  24471                                                        \\ \hline
		\end{tabular}
	\end{table}

	Overall, we performed more than 1000 experiments for the four different types of background knowledge and different perspectives. 200 different settings are used based on the following values for the main parameters: $L \in \{2,3,4,5,6\}$, $K \in \{20,30,40,50,60\}$, $C \in \{0.2,0.3,0.4,0.5\}$, and $T \in \{hours,minutes\}$. 
	We consider equal weights for the privacy gain and utility loss aspects of the score, i.e., $\alpha=0.5$ and $\beta=0.5$. 
	In Sepsis-Cases, \enquote{diagnose} and \enquote{age} are considered as the sensitive case attribute. The numerical attributes are converted to categorical attributes using \textit{boxplots} such that all the values greater than the \textit{upper quartile} are categorized as \textit{high}, the values less than the \textit{lower quartile} are categorized as \textit{low}, and the values in between are categorized as \textit{middle}. Note that the confidence value $C$ should not be greater than 0.5, i.e., there are at least two different sensitive values for a victim case. 
	To show and interpret the results of experiments, we focus on specific \textit{strong} and \textit{weak} settings. 
	We use $T=minutes$, $L=2$, $K=20$, and $C=0.5$ as the weak setting, and $T=minutes$, $L=6$, $K=60$, and $C=0.2$ as the strong setting.
	Note that in the experiments, $\tlkc$ refers to the algorithm presented in \cite{rafieitlkc} which has been extended here w.r.t. the different perspectives, i.e., Algorithm~\ref{alg:tlkc}, and $\tlkc$-$EXT$ refers to Algorithm~\ref{alg:tlkc2}.

	\subsection{Control-flow Perspective}\label{sec:exp_controlflow}
	In this subsection, we evaluate the effect of applying the extended $\tlkc$-privacy on the \textit{result utility} and \textit{data utility} with respect to the control-flow perspective. 
	We perform the control-flow perspective analysis for both event logs.

	\subsubsection{Result Utility} 
	As mentioned, for the result utility analysis of the control-flow perspective, we focus on \textit{process discovery}. The main goal is to find out \textit{how accurately the discovered process model from a privacy-aware event log capture the behavior of the original event log}. 
	To this end, we first discover a process model $M'$ from the privacy-aware event log $EL'$. Then, for $M'$, we calculate \textit{fitness}, \textit{precision}, and \textit{f1-score}, as some model quality measures, w.r.t. the original event log $EL$.
	
	\begin{figure}[t]
		\centering
		\subfloat[The measures with the weak setting.]{\includegraphics[width=0.49\textwidth]{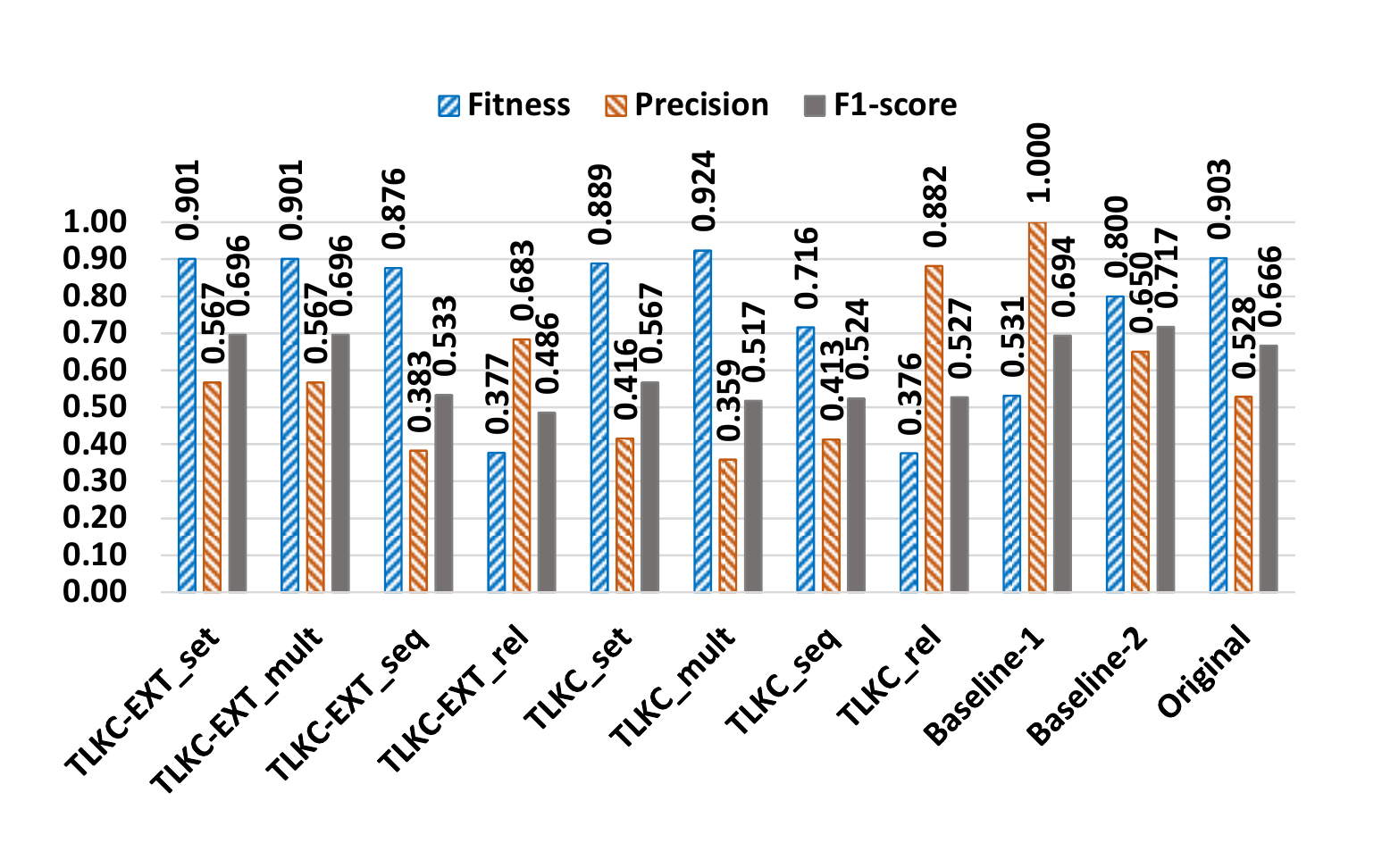}\label{fig:result-sepsis-weak}}
		\hfill
		\subfloat[The measures with the strong setting.]{\includegraphics[width=0.49\textwidth]{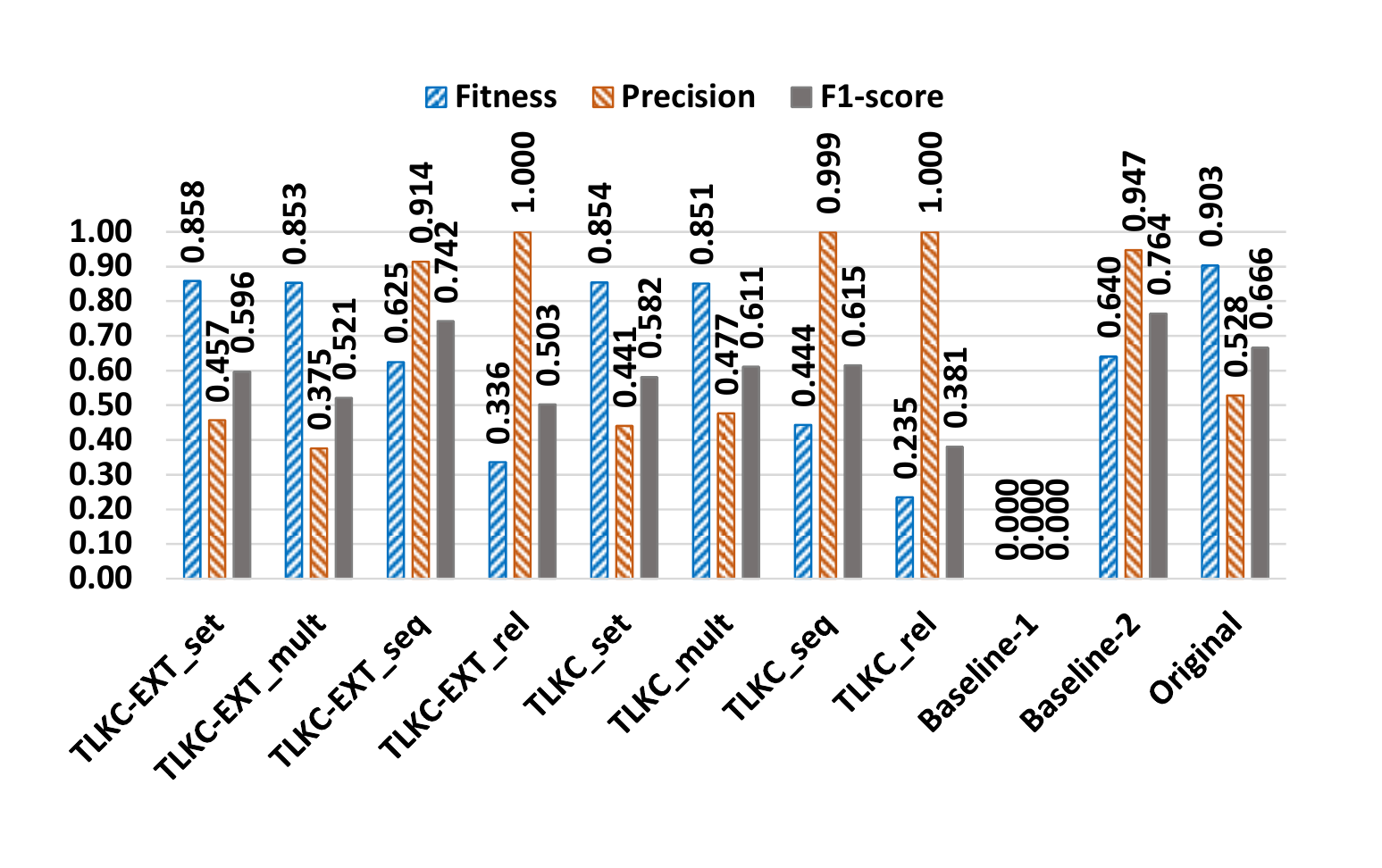}\label{fig:result-sepsis-strong}} \hfill
		\caption{The quality measures comparison between the four variants of $\tlkc$ and $\tlkc$-$EXT$, the original results, and the baseline methods for Sepsis-Cases.}
		\label{fig:discovery_eval_result}
	\end{figure}

	\textit{Fitness} quantifies the extent to which the discovered model can reproduce the traces recorded in the event log \cite{adriansyah2011conformance}.
	\textit{Precision} quantifies the fraction of the traces allowed by the model which is not seen in the event log \cite{adriansyah2015measuring}, and 
	\textit{f1-score} combines the fitness and precision ${f1\text{-}score = \nicefrac{2 \times precision \times fitness}{precision + fitness}}$.
	For process discovery, we use the \textit{inductive miner infrequent} algorithm \cite{leemans2013discovering} with the default parameters (noise threshold 0.2). 
	\autoref{fig:discovery_eval_result} shows the results of experiments for the quality measures. We consider four variants of our privacy preservation technique based on the introduced types of background knowledge where the attribute is \textit{activity}, i.e., $bk_{set,at}$, $bk_{mult,at}$, $bk_{seq,at}$, and $bk_{rel,at}$.
	Note that applying privacy preservation techniques may improve some quality measures. However, the aim is to provide as similar results as possible to the original ones and not to improve the quality of discovered models. Therefore, we include the results from the original event log to compare the proximity of the values.

	Figure~\ref{fig:result-sepsis-weak} and \ref{fig:result-sepsis-strong} show how the mentioned quality measures are affected by applying our method with the weak and strong settings (for $\tlkc$, we set $\Theta=0.5$). We compare the measures with the results from the original process model and the introduced baseline methods. If we only consider the quality measures, \textit{Baseline}-2 should be marked as the best one, since it results in better \textit{f1-score} values. However, the baseline methods remove more events from the original event log. Consequently, the corresponding privacy-aware event logs contain significantly less behavior compared to the original event log, and the resulting models have high \textit{precision} and \textit{f1-score}. The result utility analyses show that the extended version of the $\tlkc$-privacy leads to the more similar results to the original ones, specifically for the \textit{set} and \textit{multiset} types of background knowledge. However, the results obtained based on the \textit{relative} type of background knowledge have a worse \textit{fitness} value which is not surprising regarding the assumed background knowledge which is considerably strong, at the same time, difficult to achieve in reality.
	Note that the baseline methods do not protect event data against the \textit{attribute linkage} attack and provide weaker privacy guarantees.

	\begin{figure}[t]
		\centering
		\subfloat[ The data utility with the weak setting.]{\includegraphics[width=0.49\textwidth]{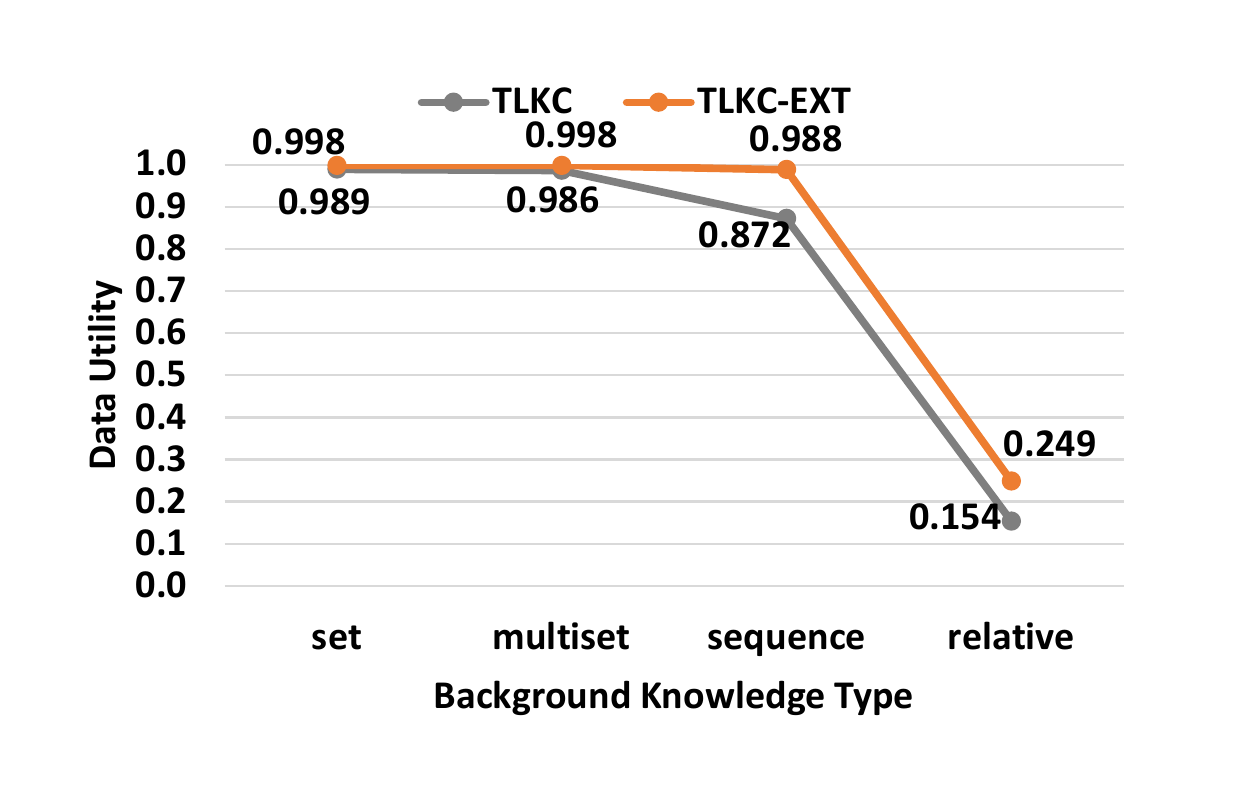}\label{fig:data-sepsis-weak}}
		\hfill
		\subfloat[ The data utility with the strong setting.]{\includegraphics[width=0.49\textwidth]{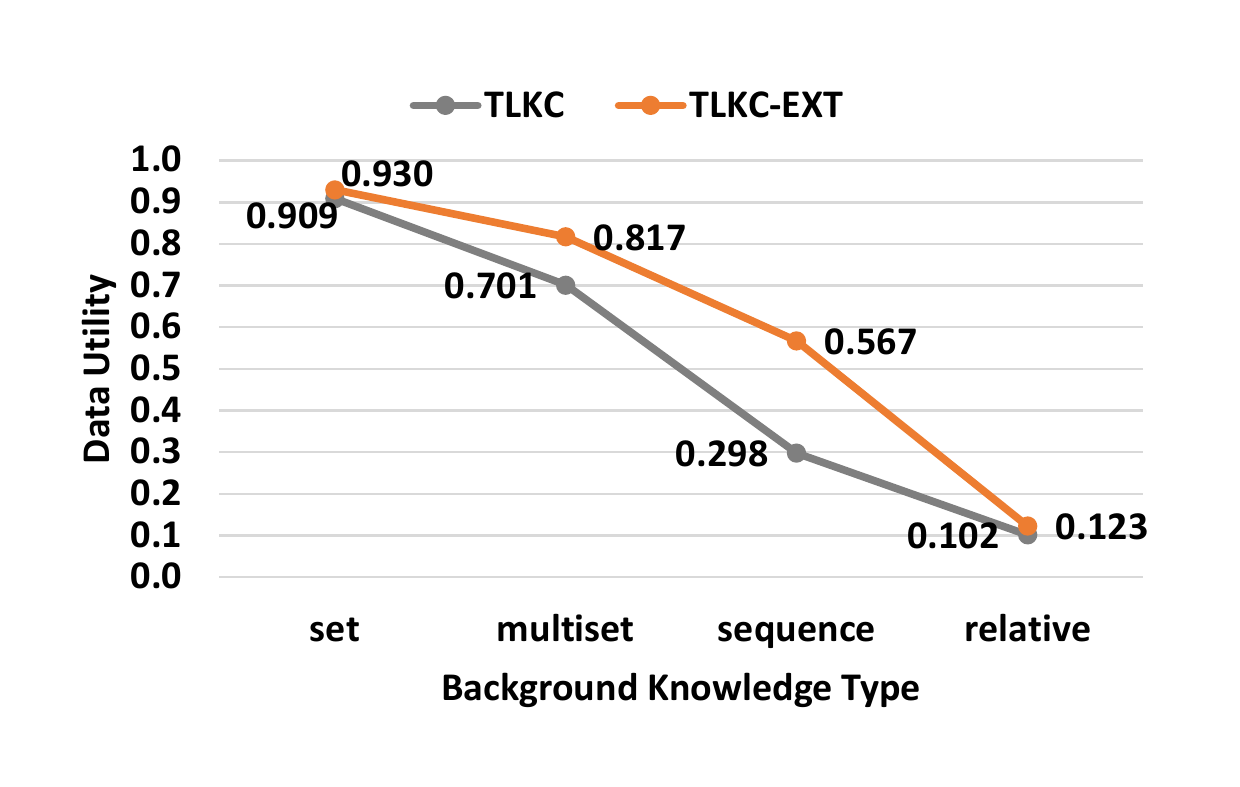}\label{fig:data-sepsis-strong}} \hfill
		\caption{The data utility comparison between $\tlkc$ and $\tlkc$-$EXT$ which provide the same privacy guarantees for the Sepsis-Cases event log.}
		\label{fig:discovery_eval_data}
	\end{figure}
	
	\begin{figure}[tb]
		\centering
		\subfloat[ The quality measures comparison between the four variants of $\tlkc$ and $\tlkc$-$EXT$, the original results, and the baseline methods.]{\includegraphics[width=0.52\textwidth]{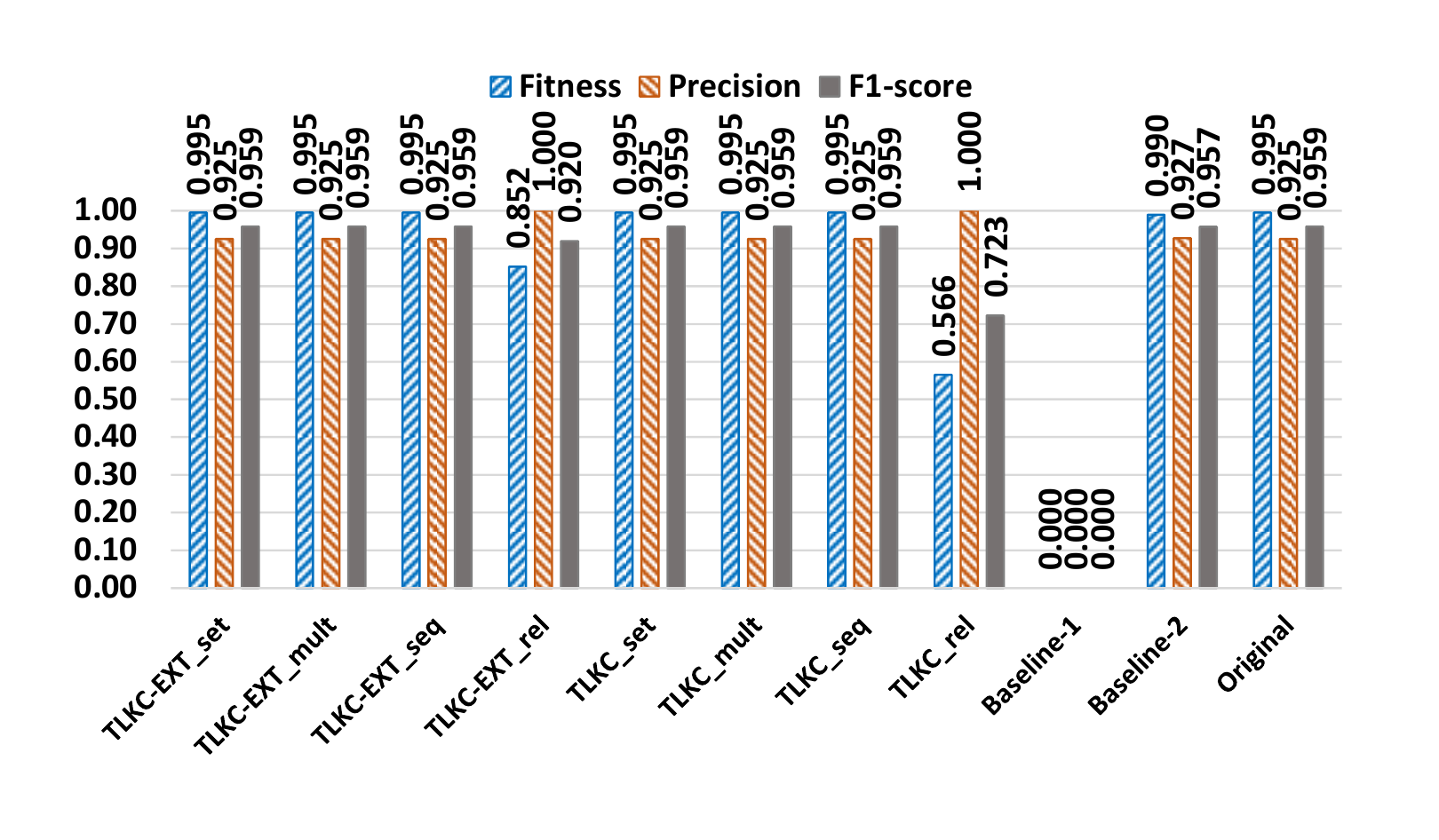}\label{fig:result-BPIC-weak}}
		\hfill
		\subfloat[ The data utility comparison between $\tlkc$ and $\tlkc$-$EXT$.]{\includegraphics[width=0.46\textwidth]{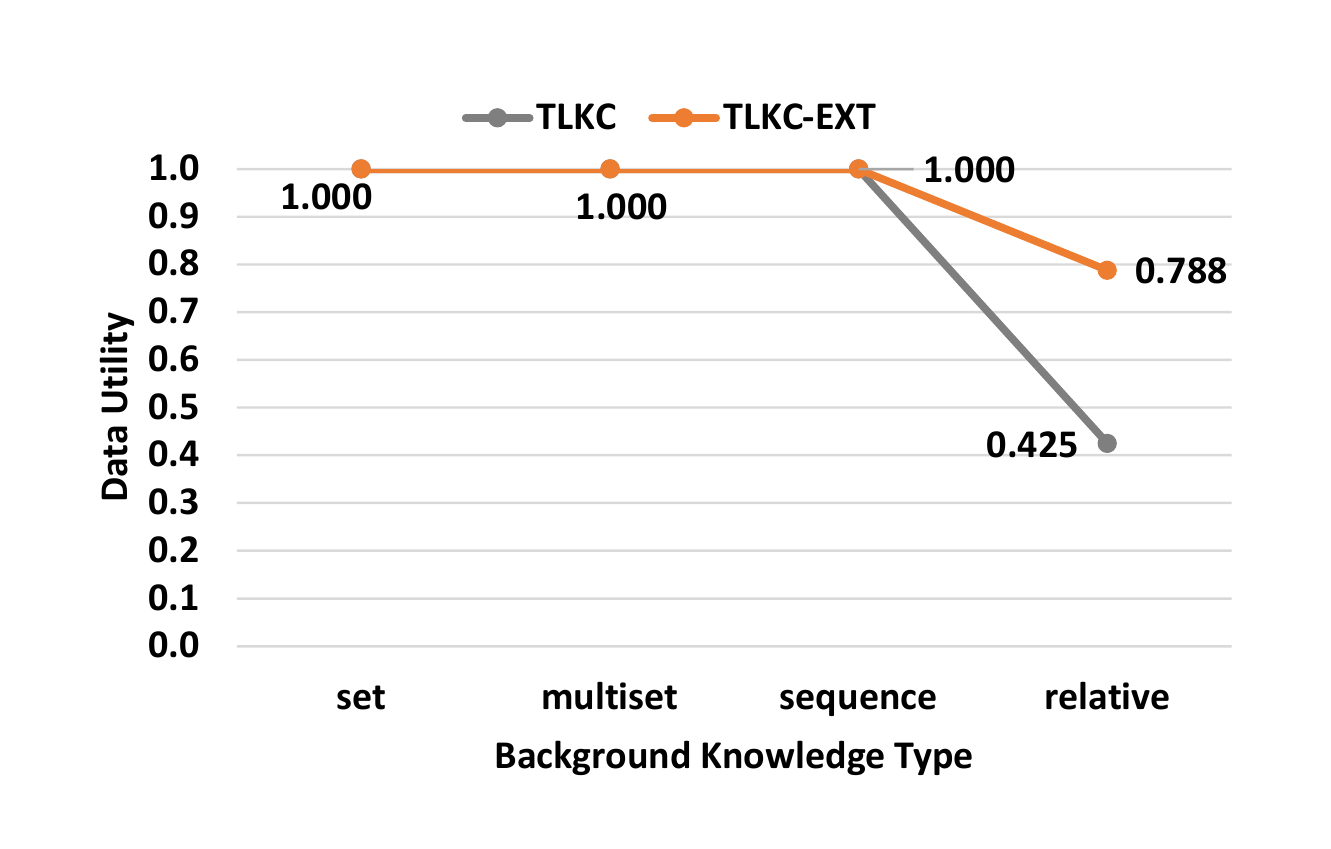}\label{fig:data-BPIC2012-strong}} \hfill
		\caption{The data and result utility analyses for BPIC-2012-APP considering the strong setting.}
		\label{fig:discovery_eval_result_data}
	\end{figure}

	\subsubsection{Data Utility}
	For the data utility analysis, we utilize the \textit{earth mover's distance}, as proposed in \cite{rafiei_quantification}. The \textit{earth mover's distance} describes the distance between two distributions \cite{emd}. In an analogy, given two piles of earth, it describes the effort required to transform one pile into the other.
	Assuming $EL$ as the original event log, $EL'$ as a privacy-aware event log, and $ps \in \perspectiveUniverse$ as the perspective of analysis. The data utility is calculated as follows: $du(EL,EL') = 1 - \min_{r \in \mathcal{RA}} ul(r,\overline{EL}_{ps},\overline{EL'}_{ps})$ where $ul(r,\overline{EL}_{ps},\overline{EL'}_{ps})$ is the distance between the traces of two event logs projected on the given perspective. Note that $r \in \mathcal{RA}$ is used as a reallocation function, and \textit{normalized edit distance} (Levenshtein) \cite{levenstein} is used to calculate the distance between variants. It should also be noted that for the control-flow $ps=\activityUniverse$.

	\autoref{fig:discovery_eval_data} shows the results of data utility analysis where we compare $\tlkc$ and $\tlkc$-$EXT$ which provide the same privacy guarantees. As can be seen, for the weak privacy setting, the data utility results are similar, and $\tlkc$-$EXT$ performs slightly better for the stronger types of background knowledge. For the strong privacy setting, $\tlkc$-$EXT$ performs considerably better for the \textit{multiset} and \textit{sequence} types of background knowledge. Comparing the data utility analysis with the result utility analysis shows that the model quality measures alone cannot precisely evaluate the effectiveness of the privacy preservation techniques. For example, in the result utility analysis, for both weak and strong setting, $\tlkc$-$EXT$ results in an acceptable \textit{f1-score} value. However, the data utility analysis shows that the utility loss is indeed high for this type of background knowledge.
	
	As already mentioned, the Sepsis-Cases event log is a significantly challenging dataset for the privacy preservation techniques due to the high uniqueness of variants. To show the effectiveness of our privacy preservation technique on other event logs, we perform the same type of analyses for BPIC-2012-APP considering only the strong setting. \autoref{fig:discovery_eval_result_data} shows that both data and result utility are high even for the strong types of background knowledge.

	\subsection{Organizational Perspective}\label{sec:exp_org}
	In this subsection, we evaluate the effect of applying the extended $\tlkc$-privacy on the \textit{result} and \textit{data} utility of the organizational perspective. The experiments of this perspective are done on BPIC-2012-APP which includes \textit{resource} information.

	\subsubsection{Result Utility}
	For the result utility analysis of organizational perspective, we focus on the \textit{social network discovery} techniques. 
	There are different methods for discovering social networks from event logs such as \textit{causality-based}, \textit{joint activities}, \textit{joint cases}, etc \cite{van2005discovering}. Here, we focus on the \textit{handover} technique which is causality-based. This technique monitors for individual cases how work moves from resource to resource, i.e., there is a \textit{handover} relation from individual $r_1$ to individual $r_2$, if there are two subsequent activities where the first is performed by $r_1$ and the second is performed by $r_2$.
	
\begin{figure}[]
		\centering
		\subfloat[ The original handover network.]{\includegraphics[width=0.40\textwidth]{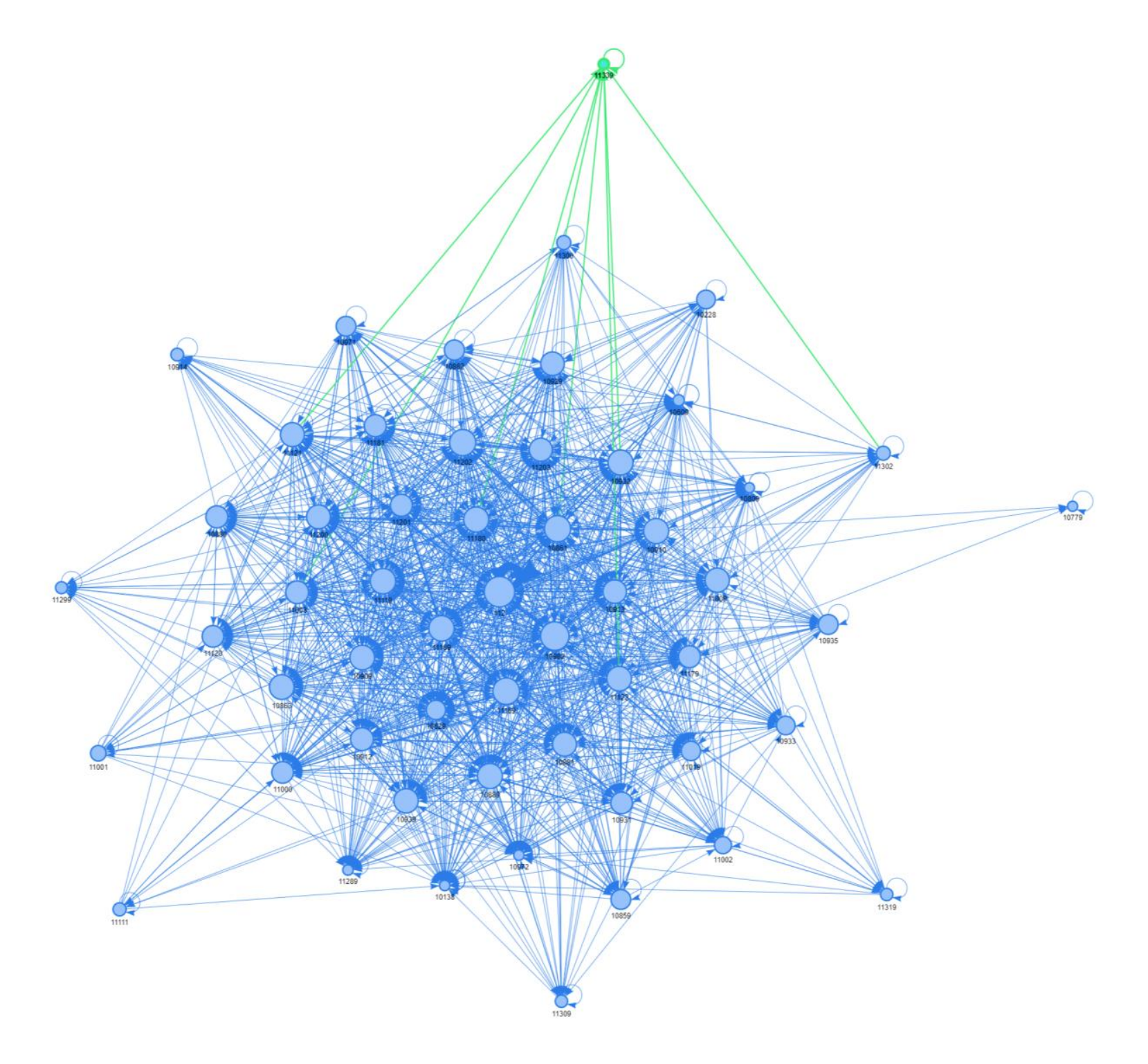}\label{fig:sn-org}}
		\hfill
		\subfloat[ The handover network discovered from the privacy-aware event log.]{\includegraphics[width=0.49\textwidth]{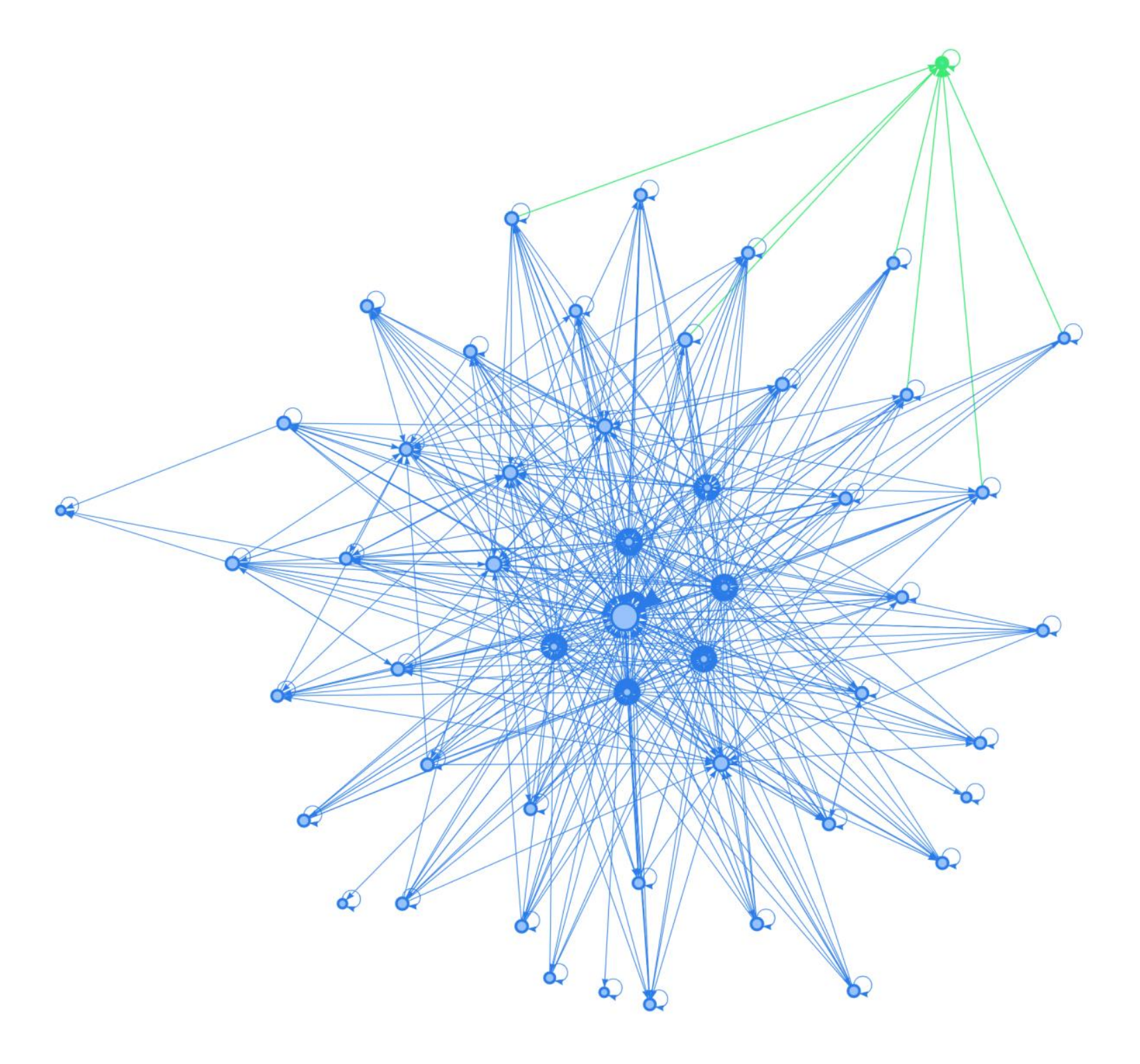}\label{fig:sn-anon}} \hfill
		\subfloat[ The relations of the resource 11339 in the original handover network. The node has 8 input links and no output link, except the self-loop.]{\includegraphics[width=0.49\textwidth]{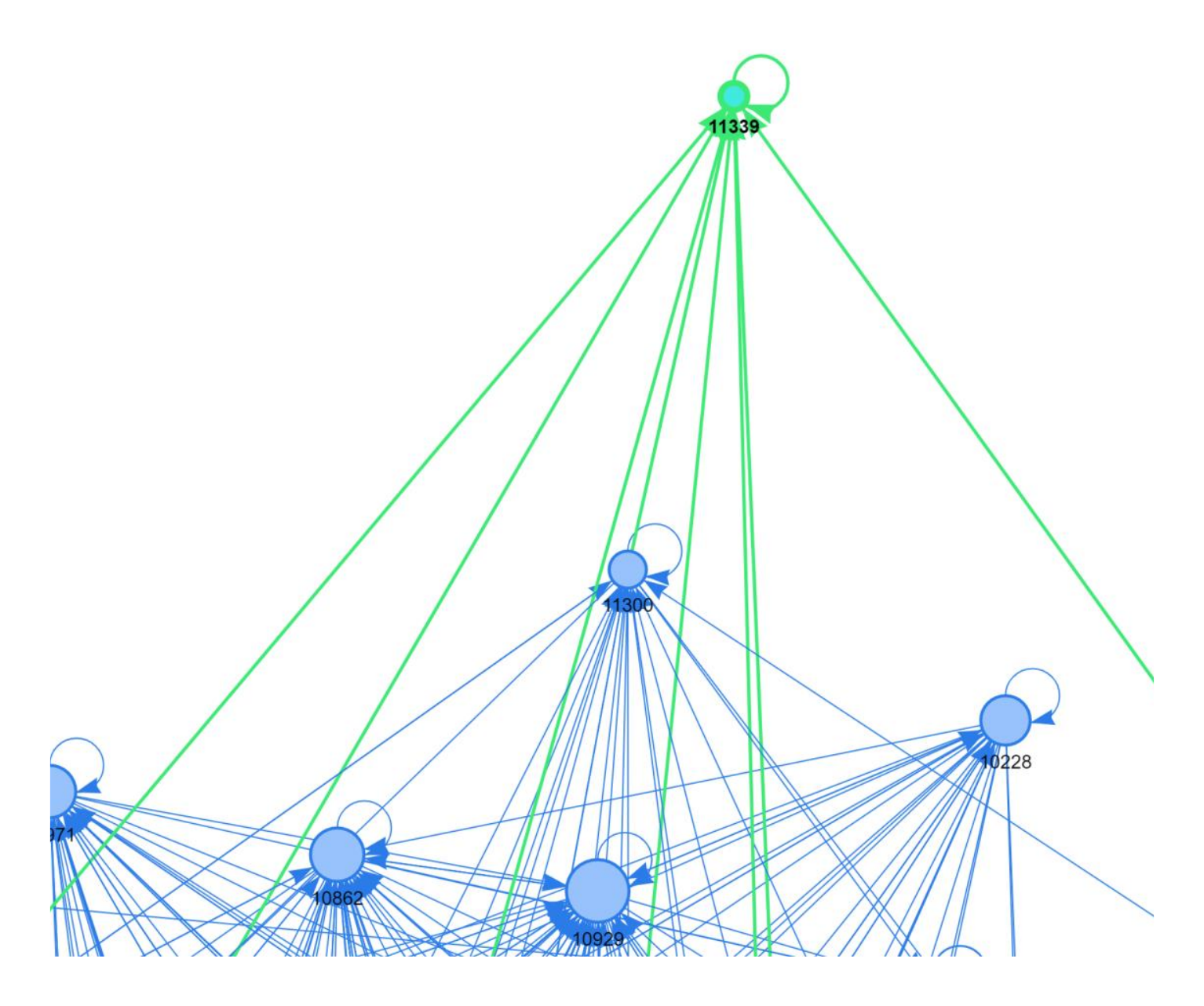}\label{fig:sn-org-zoom}} \hfill
		\subfloat[ The relations of the resource 11339 in the handover network resulting from the privacy-aware event log. The node has 7 input links and no output link, except the self-loop.]{\includegraphics[width=0.49\textwidth]{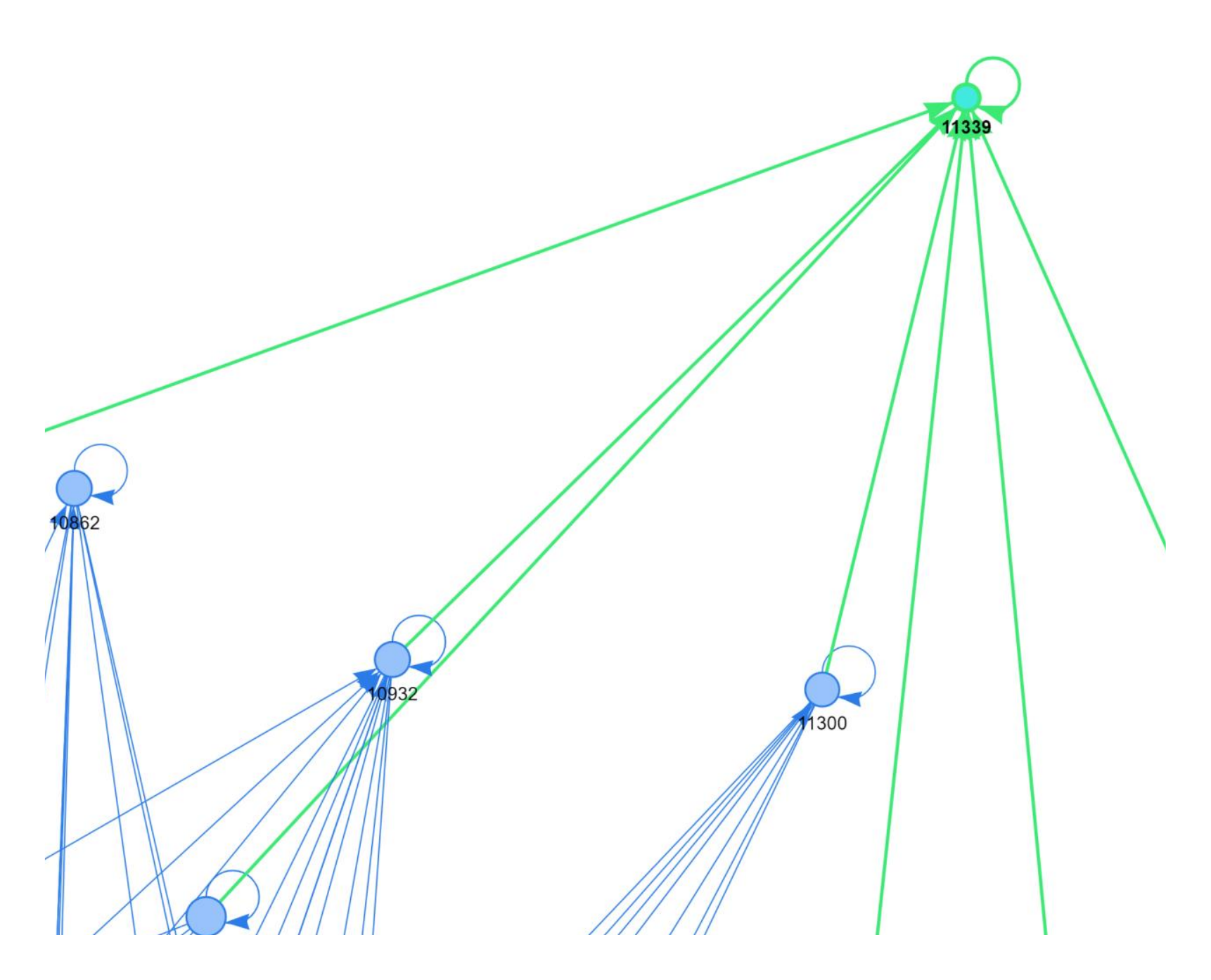}\label{fig:sn-anon-zoom}} \hfill
		\caption{The handover networks discovered from the original event log and a privacy-aware event log for BPIC-2012-APP. The privacy-aware event log was obtained using $\tlkc$-$EXT$ with the strong setting and \textit{set} as the type of background knowledge.}
		\label{fig:social-networks}
	\end{figure}

	\autoref{fig:social-networks} shows the handover networks discovered from the original event log and a privacy-aware event log when the relation threshold is 0, i.e., all the handovers. The privacy-aware event log was obtained using the $\tlkc$-$EXT$ privacy preservation technique with the strong setting and \textit{set} as the type of background knowledge. As expected, the density of the network discovered from the privacy-aware event log is less than the original handover network. However, by focusing on some specific nodes, one can see that basic concepts are preserved. For example, node 11339 in the original handover network has the following set of input links ${\{11302,11003,11300,11121,11122,11180,10932,10861\}}$ and no output link (excluding the self-loop), and in the network discovered from the privacy-aware event log, only the input link from node 11121 is removed. 
	
	\begin{figure}[t]
		\centering
		\subfloat[ The handover social network comparison for the graphs obtained from the BPIC-2012-APP event log.]{\includegraphics[width=0.49\textwidth]{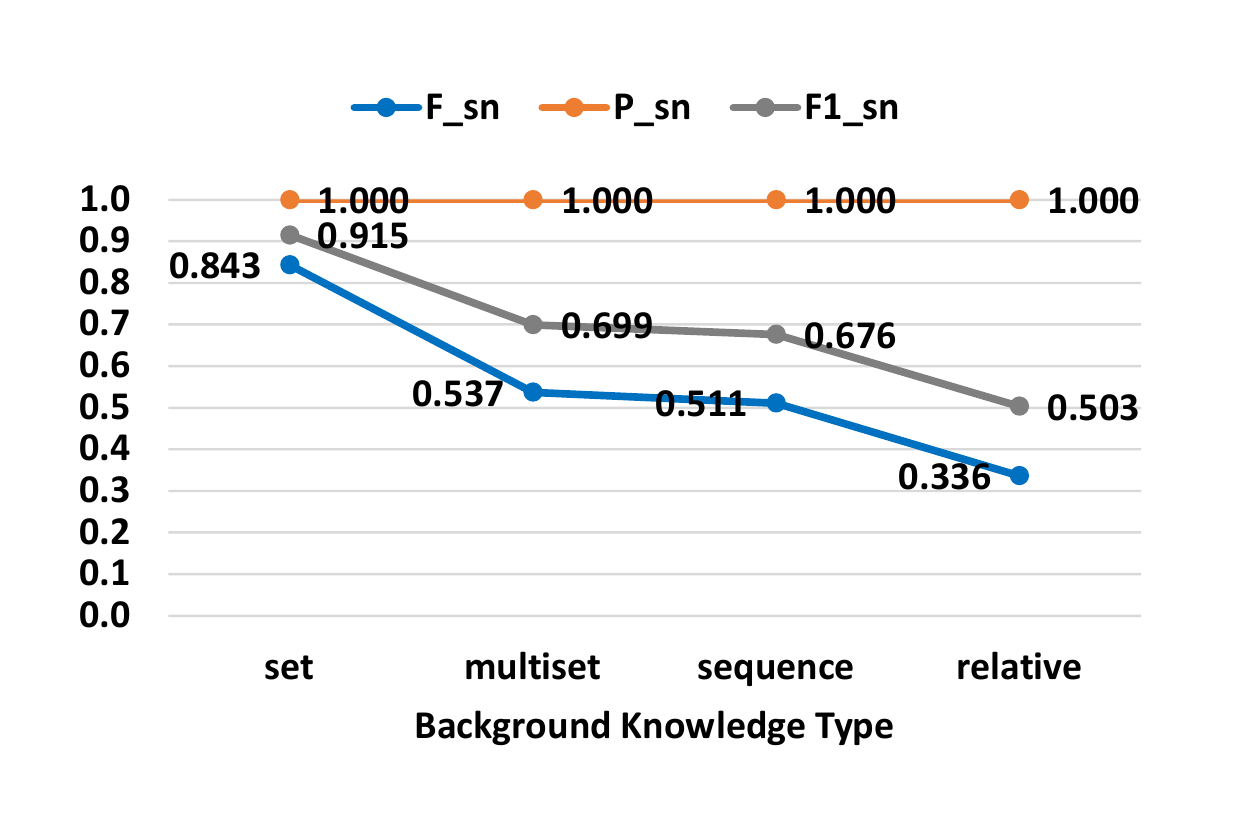}\label{fig:compare-sn-2012}}
		\hfill
		\subfloat[ The handover social network comparison for the graphs obtained from the BPIC-2017-APP event log.]{\includegraphics[width=0.49\textwidth]{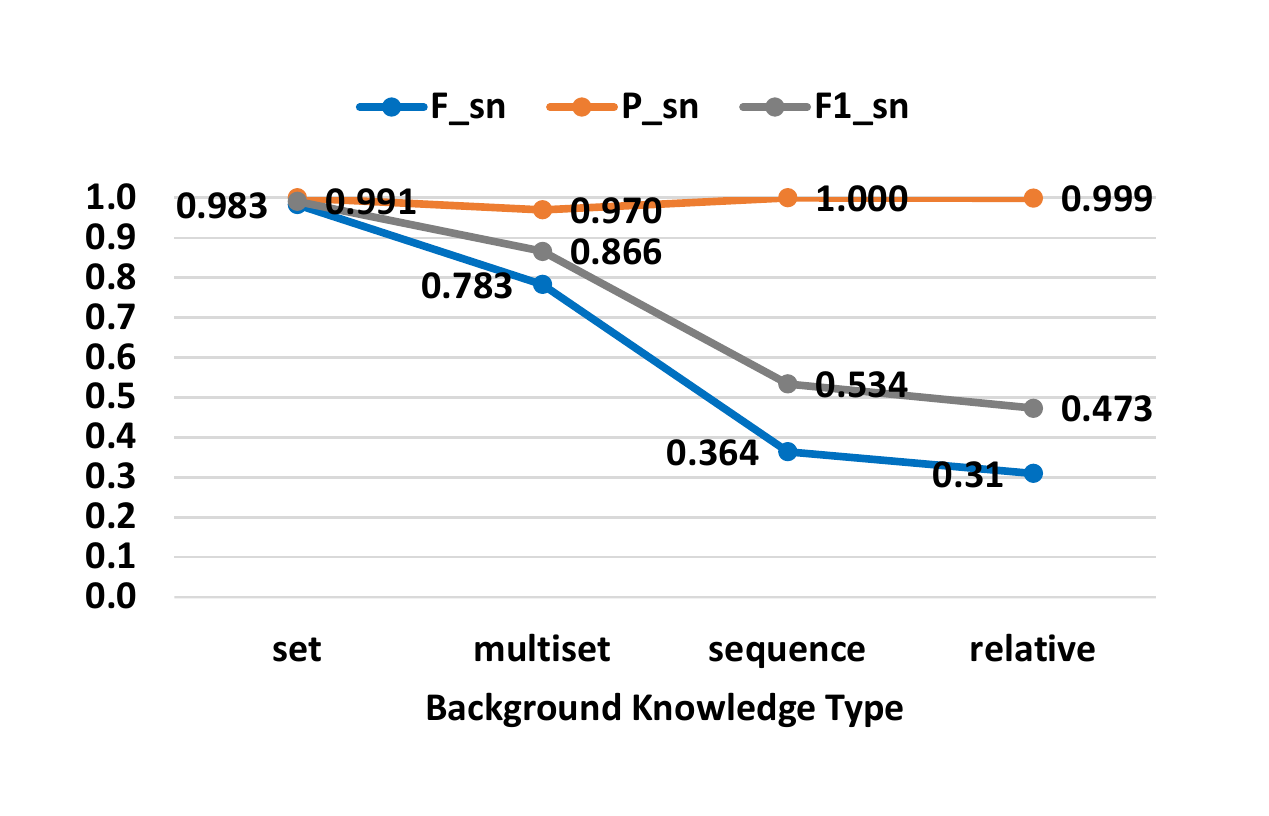}\label{fig:compare-sn-2017}} \hfill
		\caption{The social network comparison based on \textit{fitness} ($F_{sn}$), \textit{precision} ($P_{sn}$), and \textit{f1-score} ($F1_{sn}$). The privacy preservation technique is $\tlkc$-$EXT$ with the strong setting.}
		\label{fig:compare-sn}
	\end{figure}

	\begin{figure}[t]
	\centering
	\includegraphics[width=0.60\textwidth]{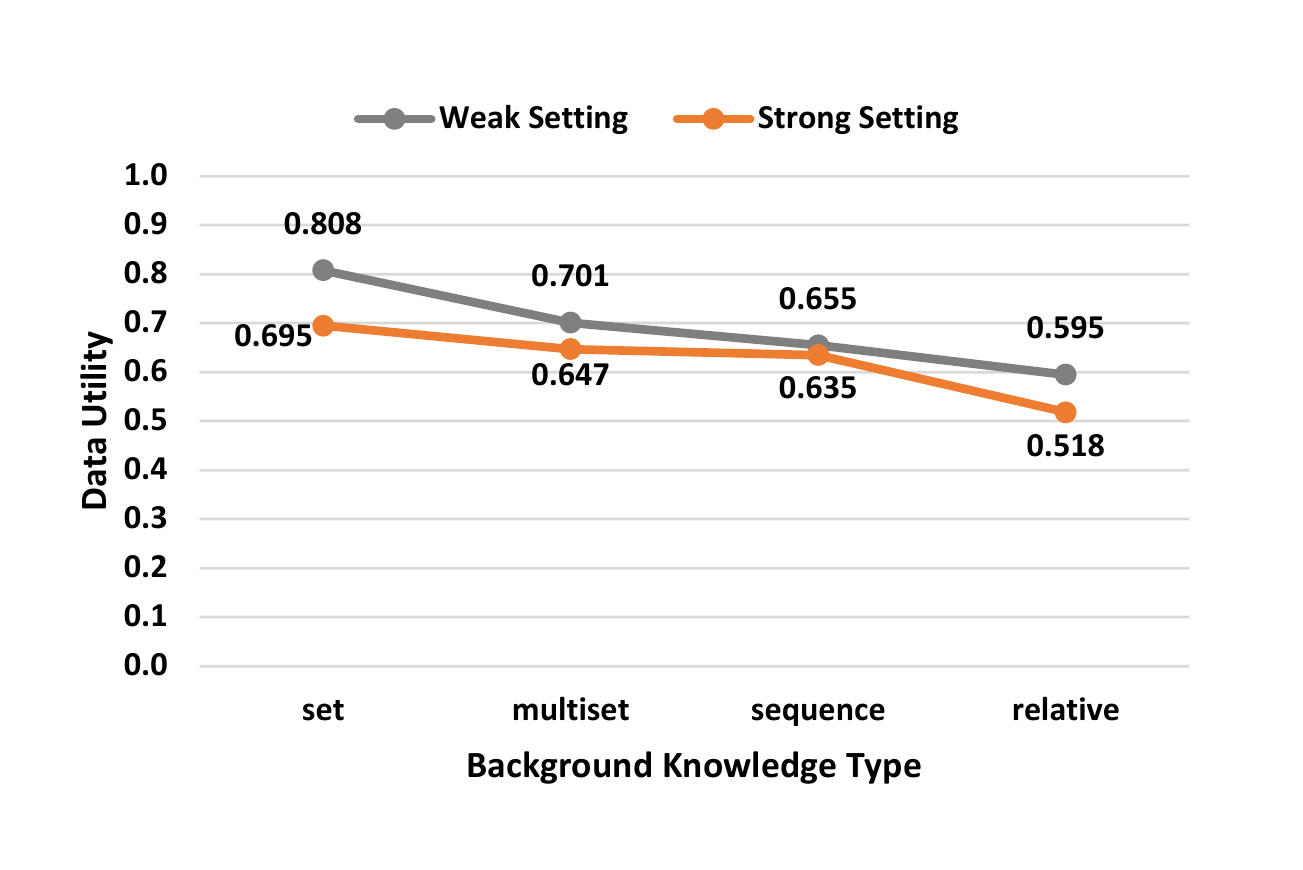}
	\caption{The data utility analysis of organizational perspective for BPIC-2012-APP with the strong and weak settings considering different types of background knowledge, and using $\tlkc$-$EXT$ as the privacy preservation technique.}\label{fig:sn-utility}
	\end{figure} 

	To quantify the similarity of social networks resulting from an original and a privacy-aware event log, we use a set of measures similar to the quality measure of process models, i.e., \textit{fitness}, \textit{precision}, and \textit{f1-score}.
	Consider $SN=(R_{EL}, DF^{EL}_{\resourceUniverse})$ and $SN'=(R_{EL'}, DF^{EL'}_{\resourceUniverse})$ as the handover social networks obtained from an original event log and its corresponding privacy-aware event log, respectively. 
	Since both $\tlkc$ and $\tlkc$-$EXT$ provide privacy guarantees by removing events, the vertices of $SN'$ is a subset of vertices in $SN$, i.e., $R_{EL} \subseteq R_{EL'}$. However, the set of edges in $SN'$ is not necessarily a subset of edges in $SN$, i.e., $SN'$ is not necessarily a subgraph of $SN$. 
	The following equations are used to compute the \textit{fitness} ($F_{sn}$) and the \textit{precision} ($P_{sn}$) for handover networks. The \textit{f1-score} for handover networks ($F1_{sn}$) is the harmonic mean of $F_{sn}$ and $P_{sn}$.

	\begin{minipage}{.36\linewidth}
		\tiny
		\[
		F_{sn} {=}  \frac{\displaystyle\sum_{(x,y) \in DF^{EL}_{\resourceUniverse} \cap DF^{EL'}_{\resourceUniverse}}|x >^{EL'}_{\resourceUniverse} y|}{\displaystyle\sum_{(x,y) \in DF^{EL}_{\resourceUniverse}} |x >^{EL}_{\resourceUniverse} y| }
		\]
	\end{minipage}%
	\begin{minipage}{.60\linewidth}
		\tiny
		\[
		P_{sn} {=}  \frac{|(R_{EL} \times R_{EL}) \setminus DF^{EL}_{\resourceUniverse} \cap (R_{EL} \times R_{EL}) \setminus DF^{EL'}_{\resourceUniverse}|}{|(R_{EL} \times R_{EL}) \setminus DF^{EL}_{\resourceUniverse}|}
		\]
	\end{minipage}

	Figure~\ref{fig:compare-sn} shows the similarity of handover social networks after applying the $\tlkc$-$EXT$ privacy model with the strong setting to BPIC-2012-APP and BPIC-2017-APP. The \textit{precision} is high for all the types of background knowledge, i.e., the handover social networks obtained from the privacy-aware event logs often do not contain edges that do not exist in the original network. The \textit{fitness} decreases when the background knowledge becomes stronger, i.e., the $SN'$s obtained based on stronger assumptions for the background knowledge have fewer edges in common with the $SN$.

	\subsubsection{Data Utility}
	For the data utility analysis of the organizational perspective, we utilize the earth mover's distance, similar to the data utility analysis of the control-flow perspective. Here, the perspective is resource, i.e., $ps = \resourceUniverse$. \autoref{fig:sn-utility} shows the results for the data utility analysis for BPIC-2012-APP considering different types of background knowledge using $\tlkc$-$EXT$ as the privacy preservation technique. As can be seen, the data utility reservation is above 0.5 even for the strong types of background knowledge.

	\subsection{Time Perspective}\label{sec:exp_time}
	We evaluate the effect on performance analyses by analyzing the bottlenecks w.r.t. the mean duration of cases between activities. Since the privacy preservation techniques may remove some activities, we cannot compare the bottlenecks in the original process model with the bottlenecks in a process model discovered from a privacy-aware event log. Therefore, we first project the original event log on the activities existing in the privacy-aware event log. Then, we discover a performance-annotated directly follows graph $DFG$ from the projected event log and compare it with the performance-annotated directly follows graph $DFG'$ from the privacy-aware event log. 
	A DFG is a graph where the nodes represent activities and the arcs represent causalities. Two activities $a_1$ and $a_2$ are connected by an arrow when $a_1$ is frequently followed by $a_2$ \cite{leemans2018scalable}.

	\begin{figure}[]
		\centering
		\subfloat[ \scriptsize $DFG'$-$bk_{set,ac}$]{\includegraphics[width=0.245\textwidth]{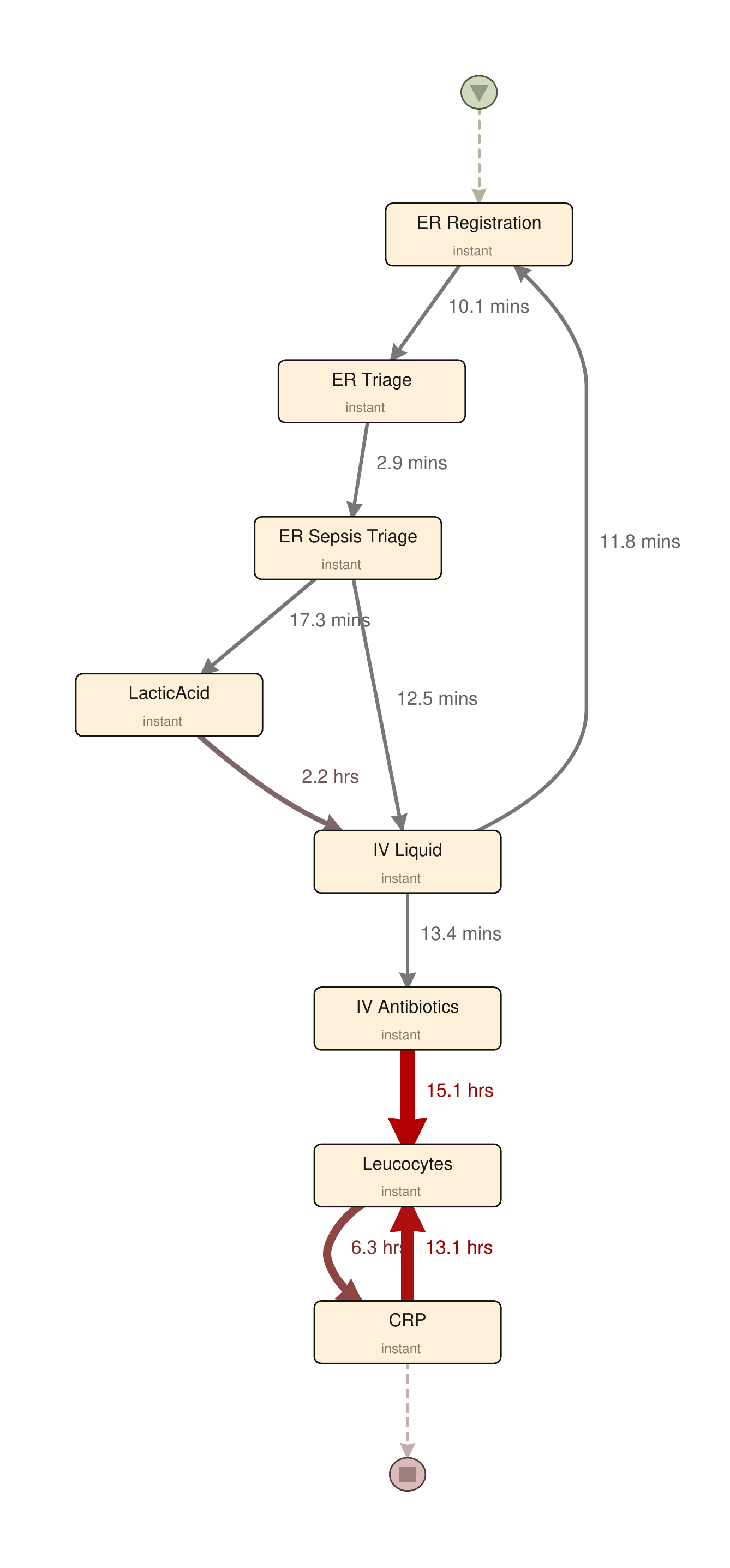}\label{fig:performance-sepsis-set-anon}}
		\subfloat[ \scriptsize $DFG$-$bk_{set,ac}$]{\includegraphics[width=0.245\textwidth]{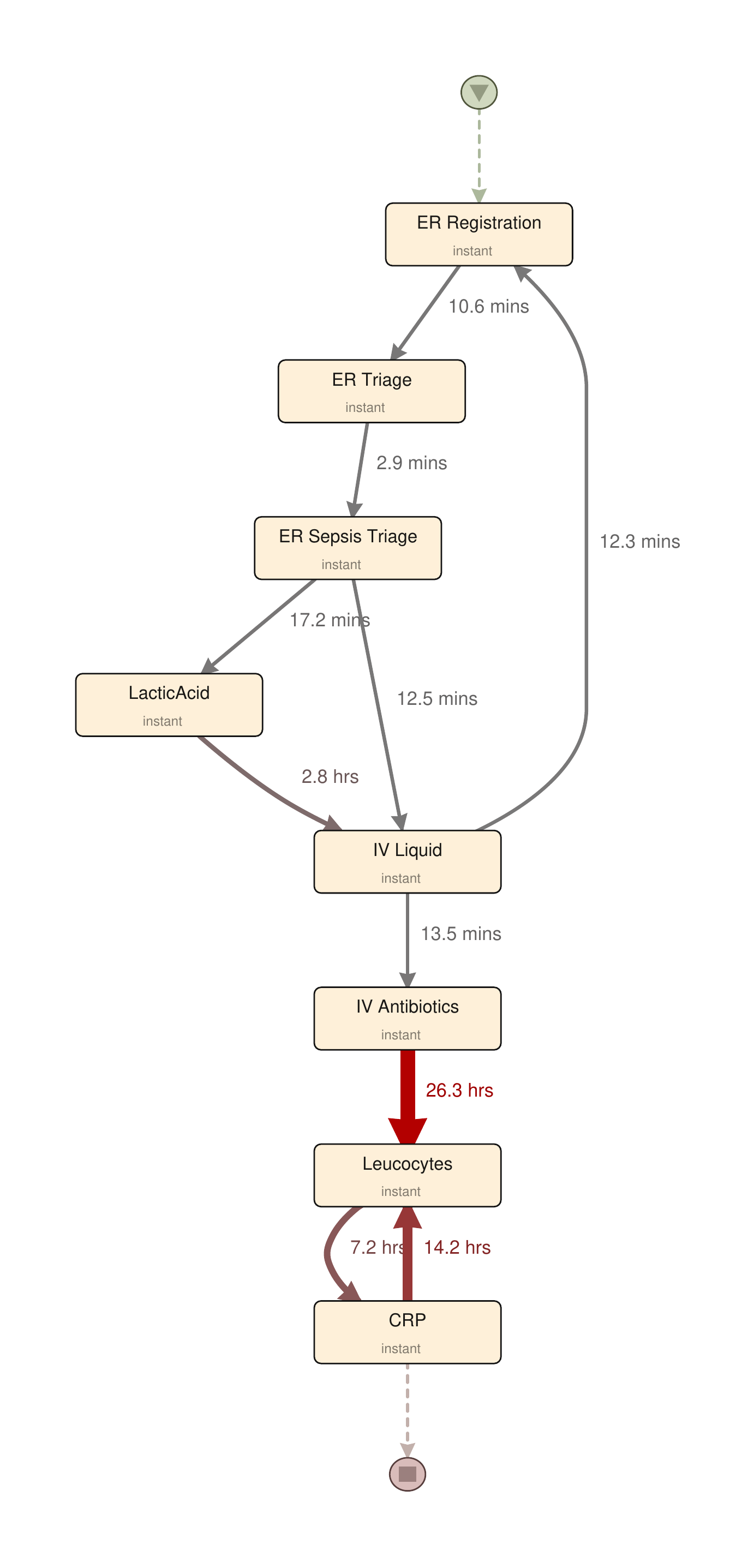}\label{fig:performance-sepsis-set-org}} 
		\label{fig:performance_set_strong}
		\subfloat[\scriptsize $DFG'$-$bk_{mult,ac}$]{\includegraphics[width=0.245\textwidth]{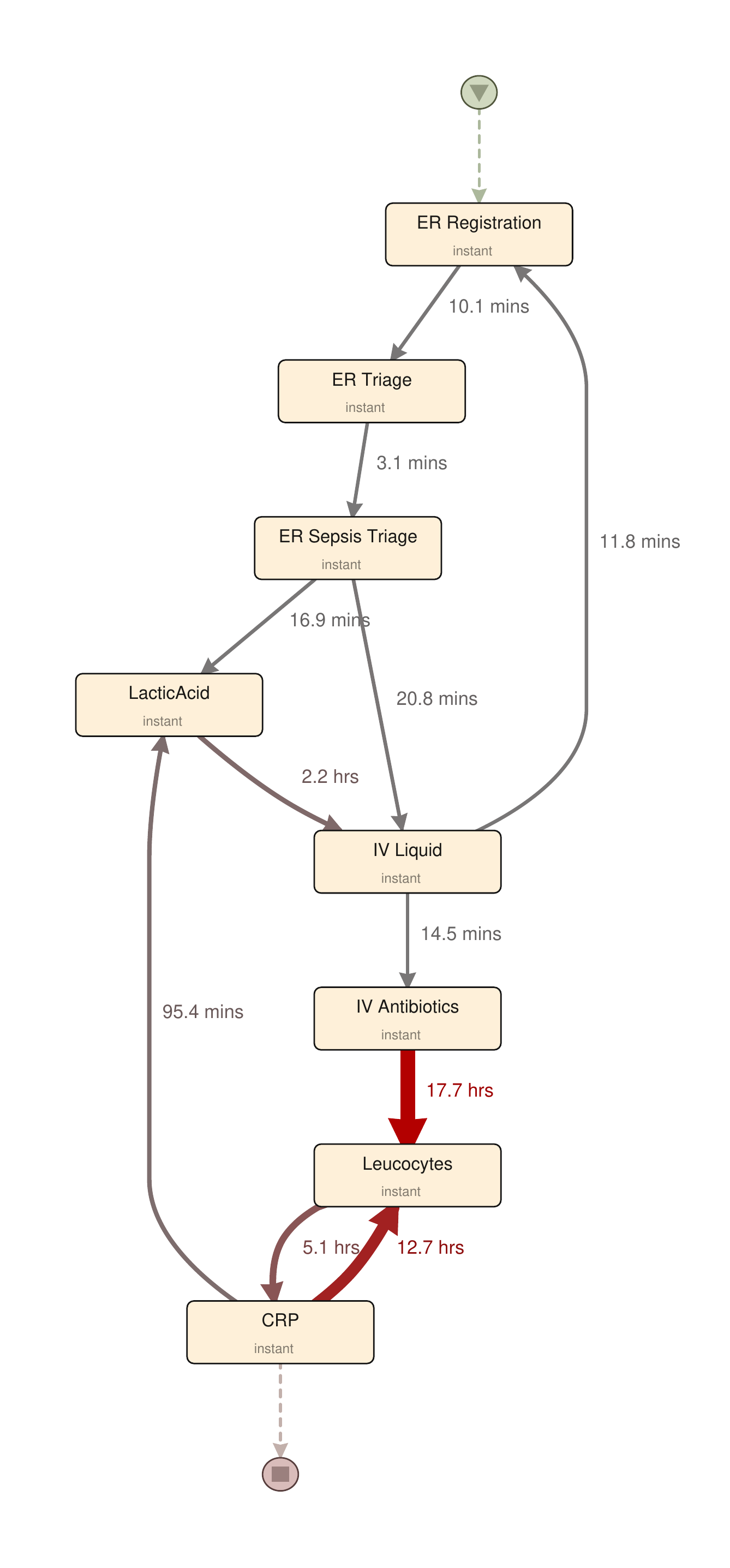}\label{fig:performance-sepsis-mult-anon}}
		\subfloat[\scriptsize $DFG$-$bk_{mult,ac}$]{\includegraphics[width=0.245\textwidth]{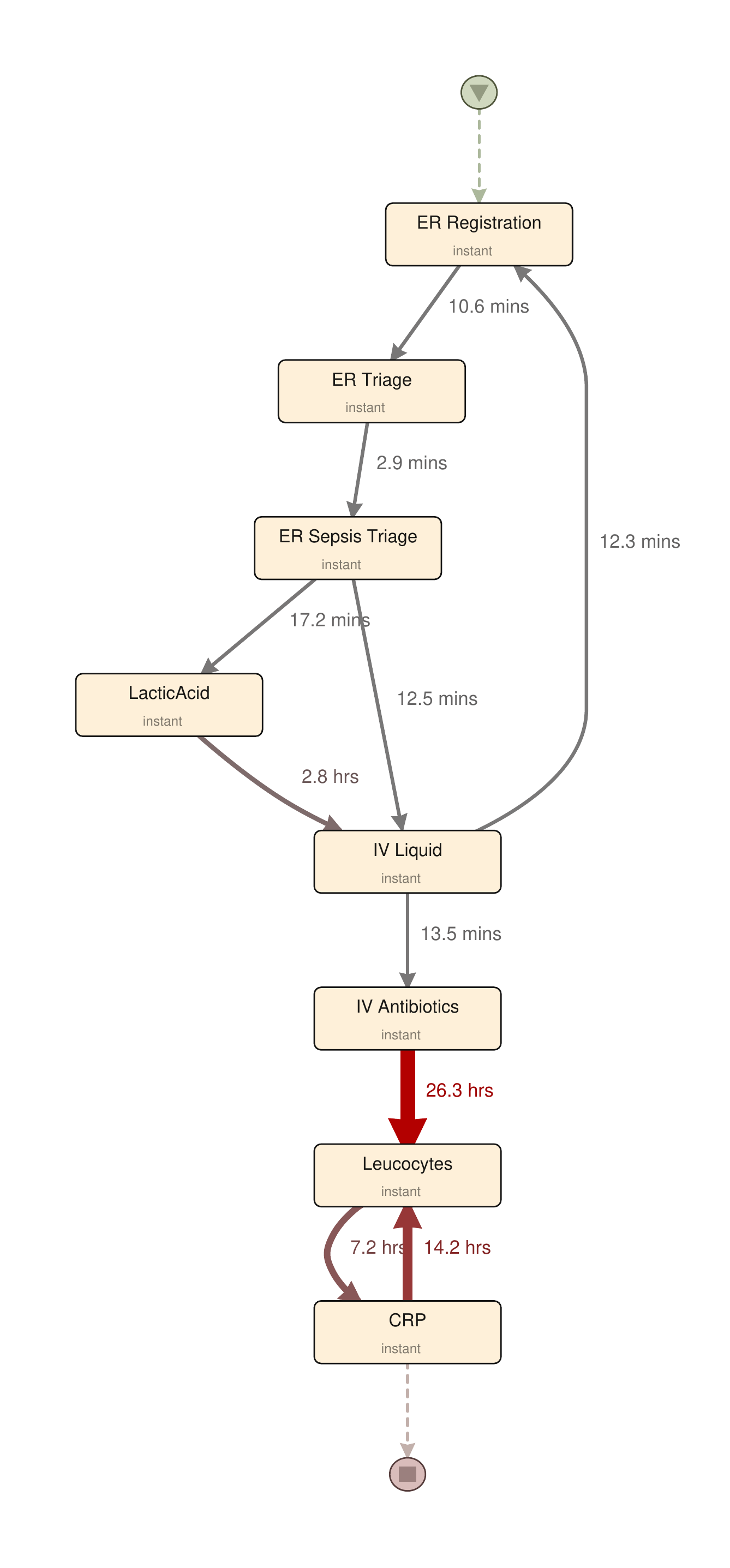}\label{fig:performance-sepsis-mult-org}} 
		\caption{The performance-annotated DFGs from the projected event log ($DFG$) and an anonymized event log ($DFG'$) for Sepsis-Cases using $\tlkc$-$EXT$ with the strong setting and the specified types of background knowledge.}
		\label{fig:performance_set_multi_strong}
	\end{figure}

	\begin{figure}[]
		\centering
		\subfloat[ \scriptsize $DFG'$-$bk_{seq,ac}$]{\includegraphics[width=0.245\textwidth]{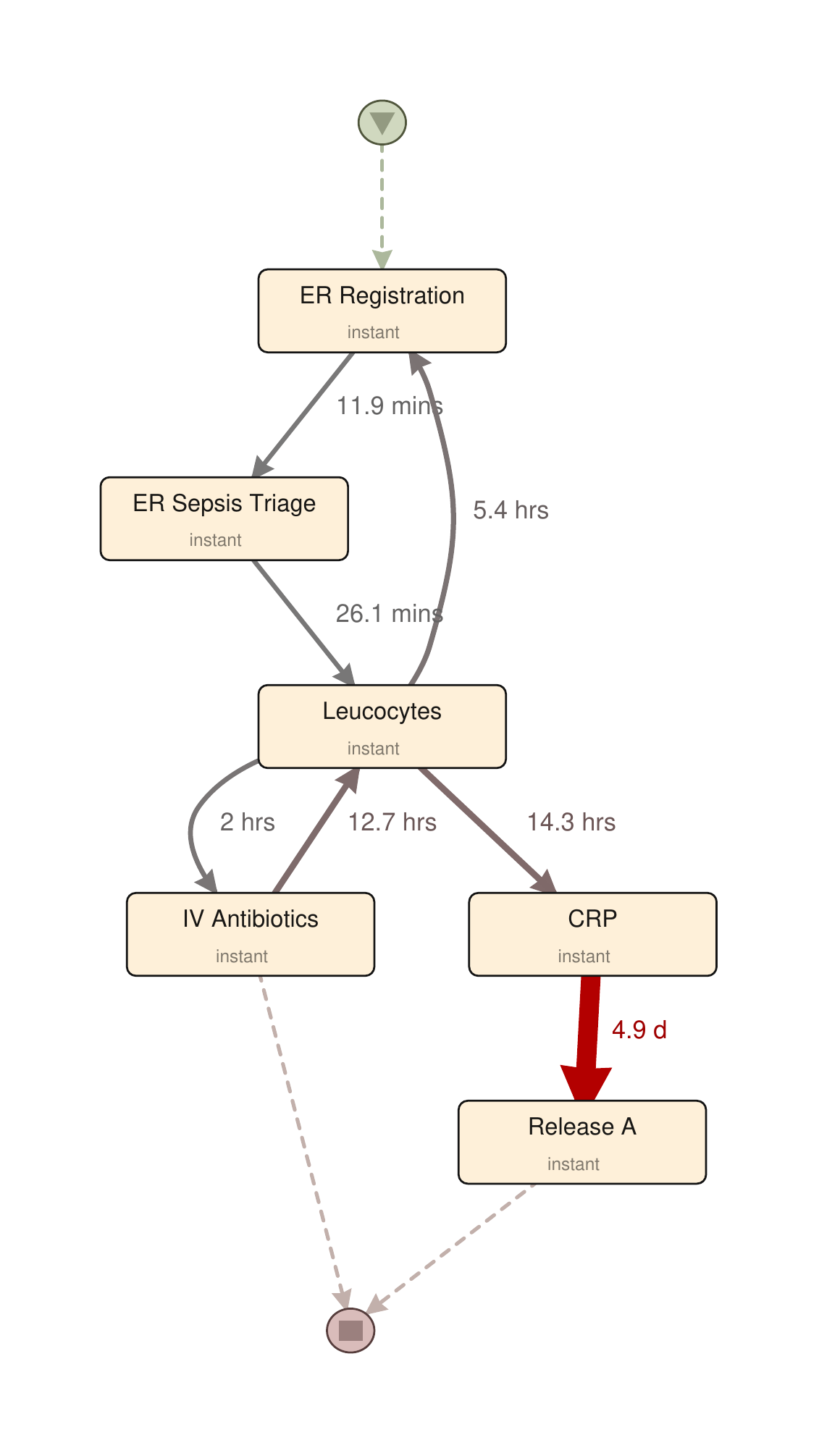}\label{fig:performance-sepsis-seq-anon}}
		\subfloat[ \scriptsize $DFG$-$bk_{seq,ac}$]{\includegraphics[width=0.245\textwidth]{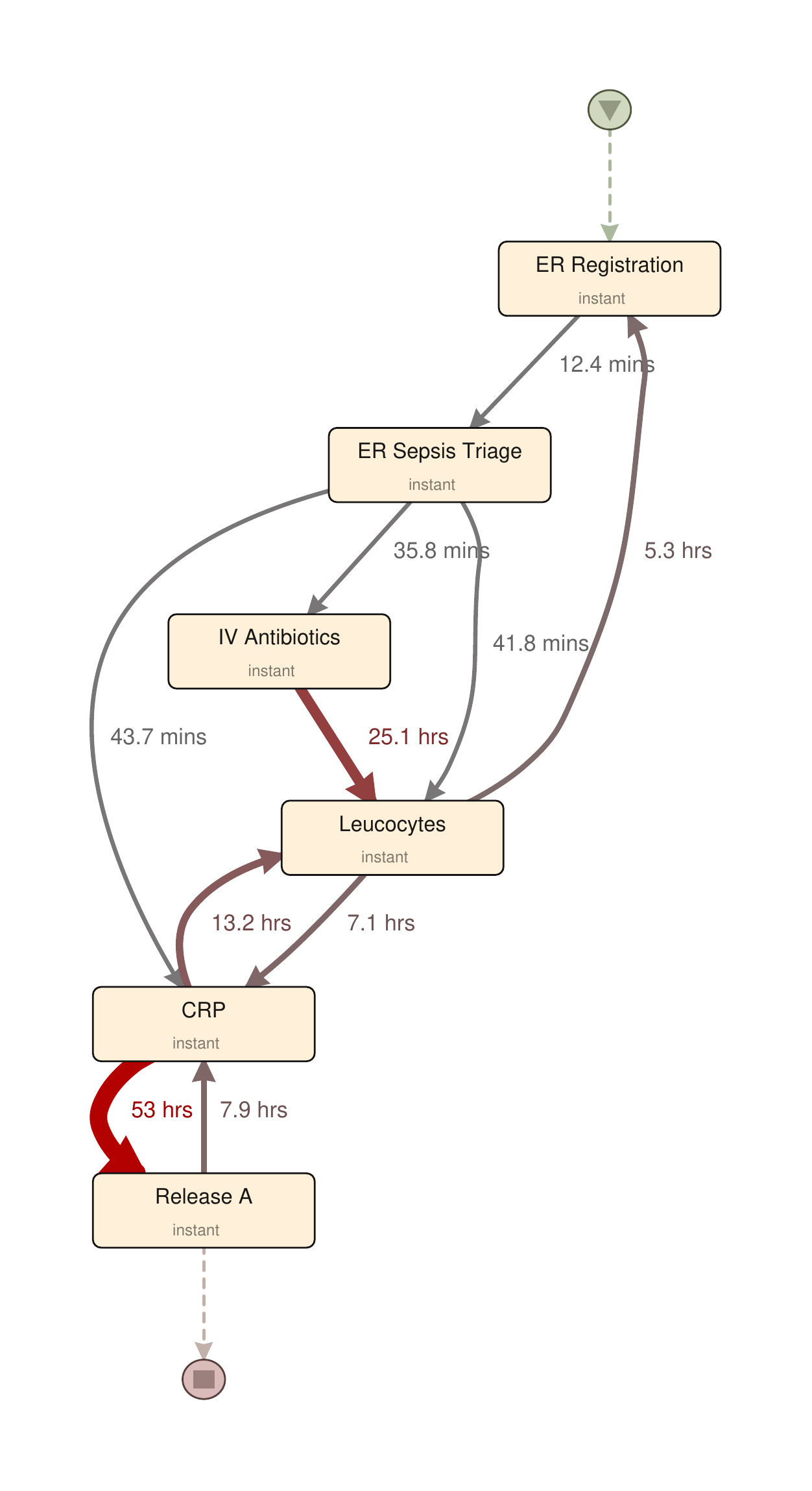}\label{fig:performance-sepsis-seq-org}} 
		\label{fig:performance_seq_strong}
		\subfloat[\scriptsize $DFG'$-$bk_{rel,ac}$]{\includegraphics[width=0.245\textwidth]{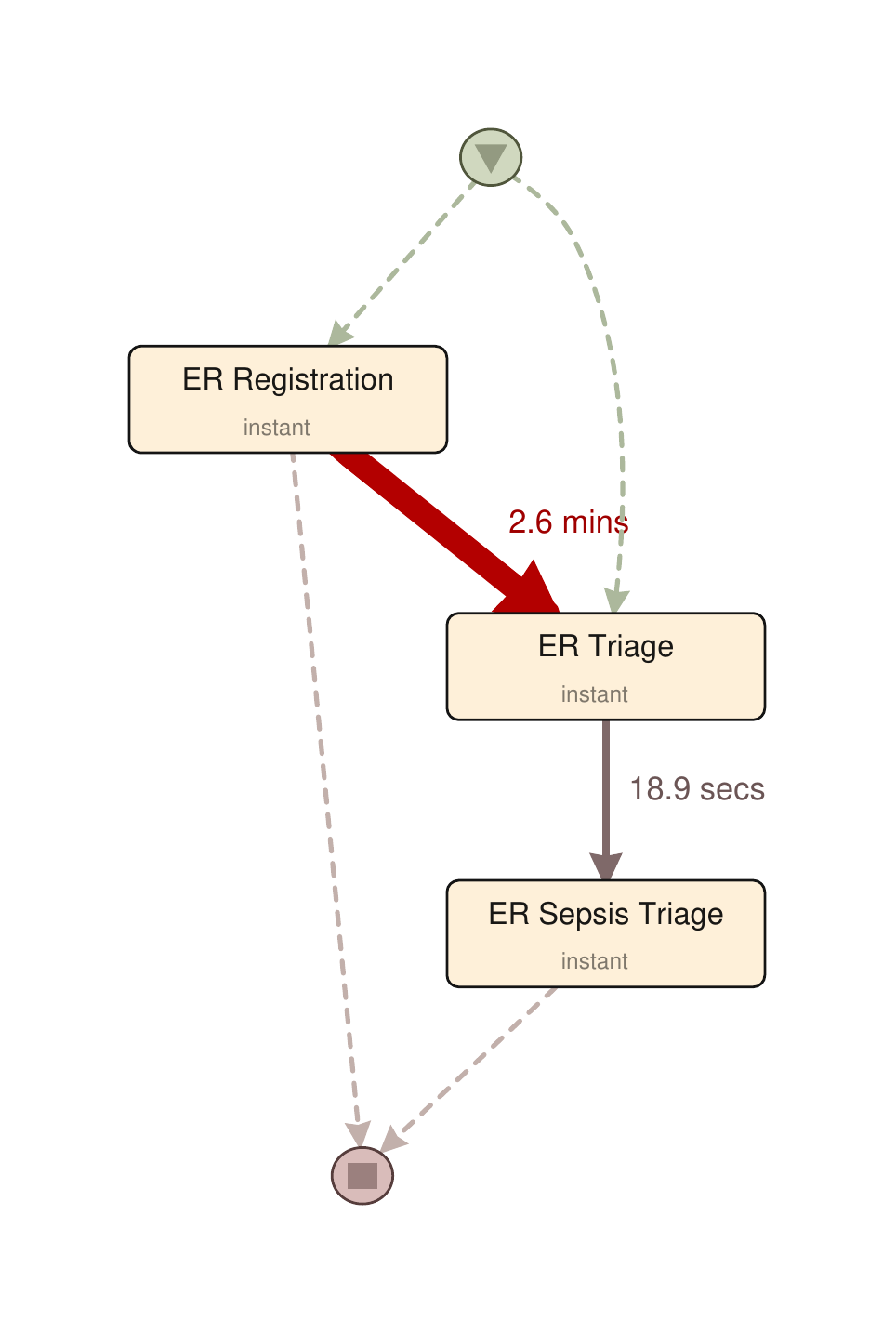}\label{fig:performance-sepsis-rel-anon}}
		\subfloat[\scriptsize $DFG$-$bk_{rel,ac}$]{\includegraphics[width=0.245\textwidth]{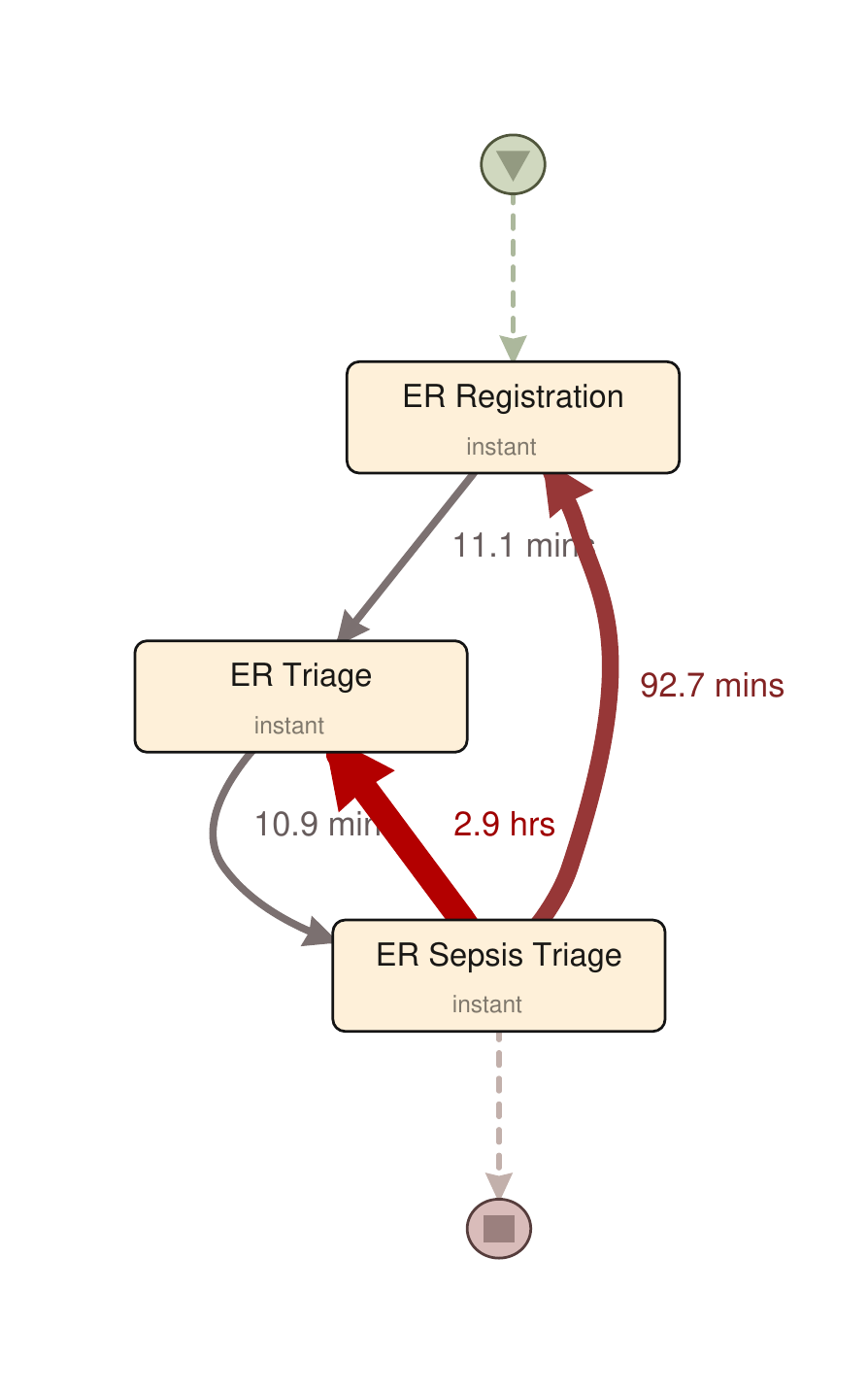}\label{fig:performance-sepsis-rel-org}} 
		\label{fig:performance_multiset_strong}
		\caption{The performance-annotated DFGs from the projected event log ($DFG$) and an anonymized event log ($DFG'$) for Sepsis-Cases using $\tlkc$-$EXT$ with the strong setting and the specified types of background knowledge.}
		\label{fig:performance_seq_rel_strong}
	\end{figure}

	\autoref{fig:performance_set_multi_strong} (\textit{set} and \textit{multiset} as the types of background knowledge) and \autoref{fig:performance_seq_rel_strong} (\textit{sequence} and \textit{relative} as the types of background knowledge) show the results for Sepsis-Cases using $TLKC$-$EXT$ with the strong setting.\footnote{The results provided by Disco (https://fluxicon.com/disco/) with the sliders set to the maximal number of activities and the minimal paths.} As can be seen, the bottlenecks in $DFG$ and $DFG'$ are the same for all the variants, except for DFGs discovered using $bk_{rel,ac}$, where the assumed background knowledge is \textit{relative} which is significantly strong and our data utility analysis in Section~\ref{sec:exp_controlflow} demonstrated a low data utility preservation for Sepsis-Cases. Note that the mean duration of the cases are different in $DFG$ and $DFG'$ due to the relative timestamps in the privacy-aware event logs.

	We also evaluate the similarity of the Directly Follows Graphs (DFGs) resulting from an original event log and its corresponding privacy-aware event log. Let $DFG{=}(A_{EL},DF^{EL}_{\activityUniverse})$ and $DFG'{=}(A_{EL'},DF^{EL'}_{\activityUniverse})$ be the directly follows graphs obtained from an original and its corresponding privacy-aware event logs, respectively.
	To compare these graphs, we follow the same approach taken for quantifying the similarity of social networks. The \textit{fitness} ($F_{dfg}$) and \textit{precision} ($P_{dfg}$) for DFGs are calculated as follows:

	\begin{minipage}{.36\linewidth}
		\tiny
		\[
		F_{dfg} {=}  \frac{\displaystyle\sum_{(x,y) \in DF^{EL}_{\activityUniverse} \cap DF^{EL'}_{\activityUniverse}}|x >^{EL'}_{\activityUniverse} y|}{\displaystyle\sum_{(x,y) \in DF^{EL}_{\activityUniverse}} |x >^{EL}_{\activityUniverse} y| }
		\]
	\end{minipage}%
	\begin{minipage}{.60\linewidth}
		\tiny
		\[
		P_{dfg} {=}  \frac{|(A_{EL} \times A_{EL}) \setminus DF^{EL}_{\activityUniverse} \cap (A_{EL} \times A_{EL}) \setminus DF^{EL'}_{\activityUniverse}|}{|(A_{EL} \times A_{EL}) \setminus DF^{EL}_{\activityUniverse}|}
		\]
	\end{minipage}

\begin{figure}[t]
		\centering
		\subfloat[ The DFG comparison for the graphs obtained from the Sepsis-Cases event log.]{\includegraphics[width=0.49\textwidth]{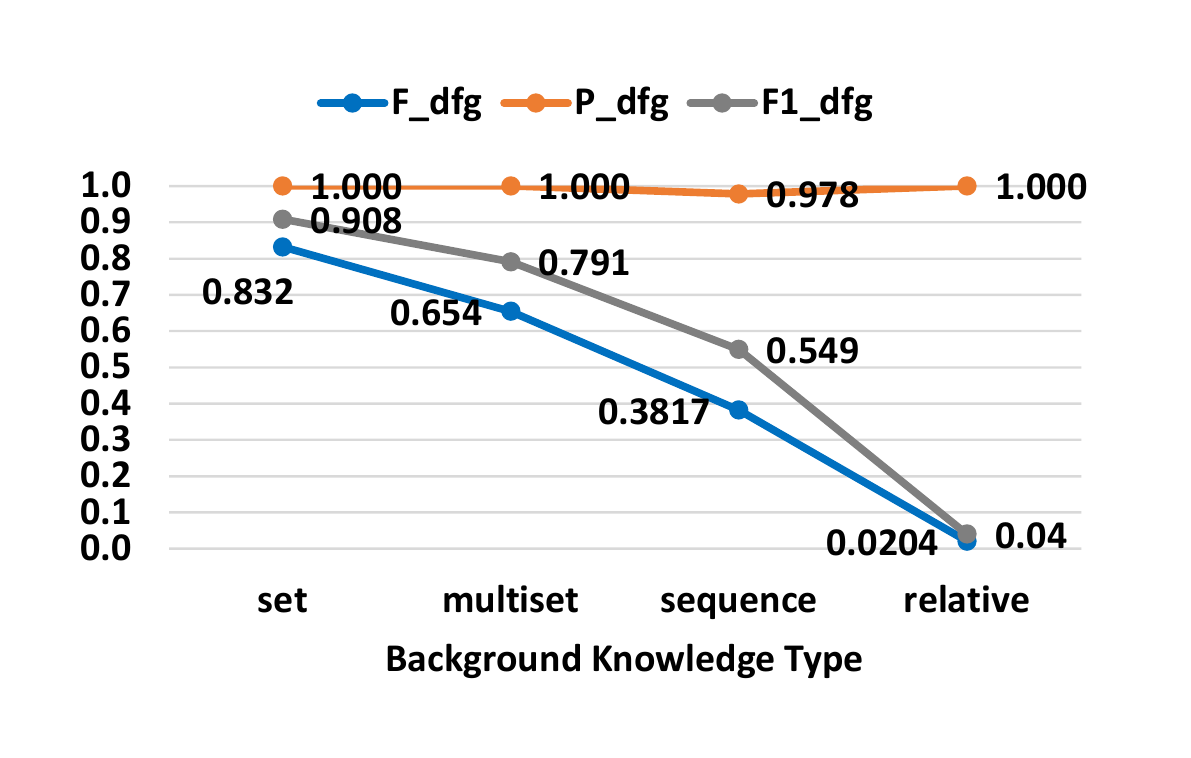}\label{fig:comapre-dfg-sepsis}}
		\hfill
		\subfloat[ The DFG comparison for the graphs obtained from the BPIC-2012-APP event log.]{\includegraphics[width=0.49\textwidth]{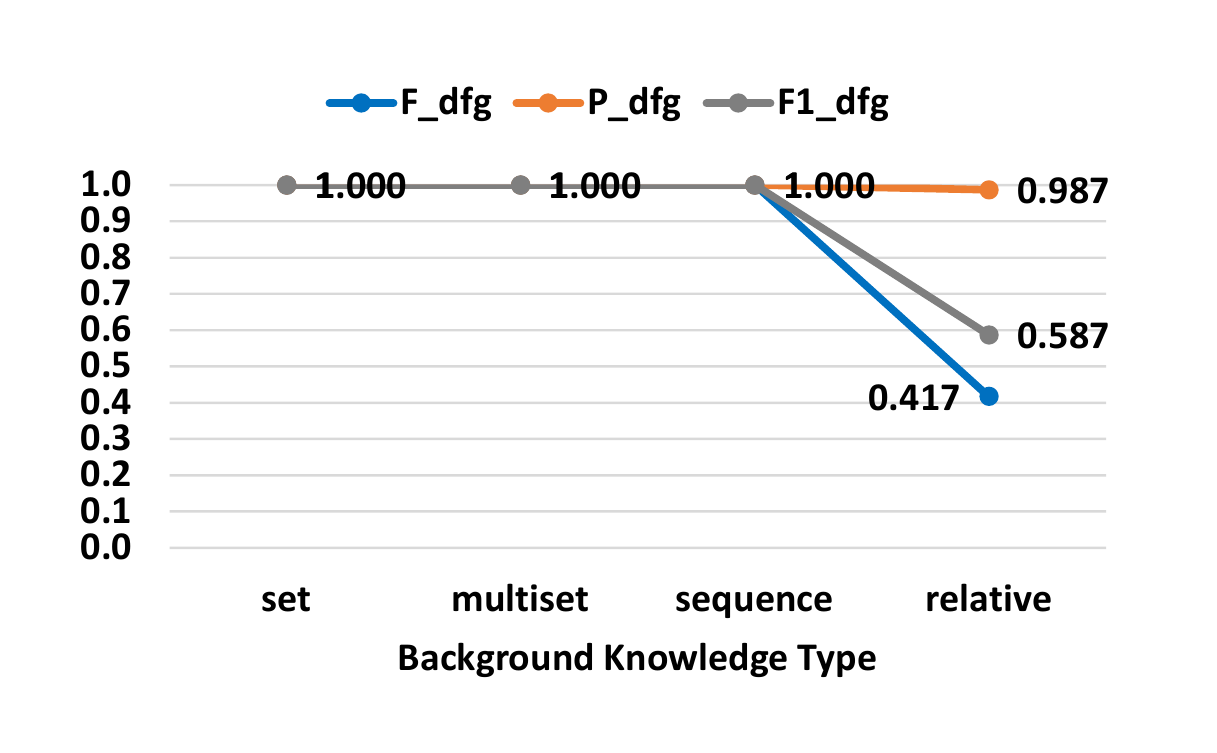}\label{fig:compare-dfg-2012}}
		\hfill
		\subfloat[ The DFG comparison for the graphs obtained from the BPIC-2017-APP event log.]{\includegraphics[width=0.49\textwidth]{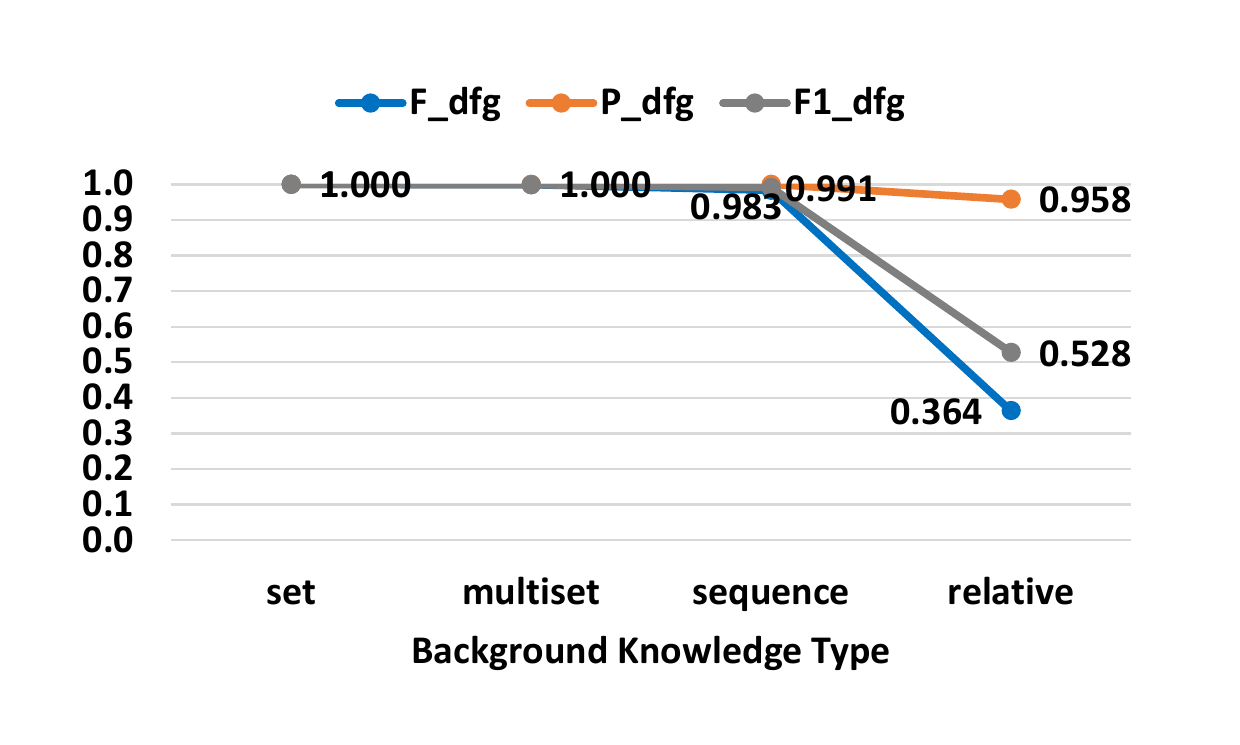}\label{fig:comapre-dfg-2017}} \hfill
		\caption{The DFG comparison based on \textit{fitness} ($F_{dfg}$), \textit{precision} ($P_{dfg}$), and \textit{f1-score} ($F1_{dfg}$). The privacy preservation technique is $\tlkc$-$EXT$ with the strong setting.}
		\label{fig:compare-dfg}
	\end{figure}

	The \textit{f1-score} for DFGs ($F1_{dfg}$) is the harmonic mean of $F_{dfg}$ and $P_{dfg}$. Figure~\ref{fig:compare-dfg} shows the similarity of DFGs after applying the $\tlkc$-$EXT$ privacy model with the strong setting for Sepsis-Cases, BPIC-2012-APP, and BPIC-2017-APP.
	The \textit{precision} is always high, i.e., the DFGs obtained from the privacy-aware event logs often do not contain directly follows relations that do not exist in the original DFG. For the Sepsis-Cases event log, the \textit{fitness} decreases when the background knowledge becomes stronger, i.e., the $DFG'$s obtained based on stronger assumptions for the background knowledge preserve fewer directly follows relations of the original DFG. The \textit{fitness} for the BPIC event logs only drops for the relative type of background knowledge which is considerably strong.

	\section{Related Work}\label{sec:relatedwork}
	In process mining, the research field of confidentiality and privacy is recently receiving more attention. In this section, we list the work that has been done in this research field which is rapidly growing. 
	
	In \cite{van2016responsible}, \textit{Responsible Process Mining} (RPM) is introduced as the sub-discipline which focuses on possible negative side-effects of applying process mining where \textit{Fairness}, \textit{Accuracy}, \textit{Confidentiality}, and \textit{Transparency} (FACT) are considered as the concerns. 
	In \cite{mannhardt2018privacy}, the authors provide an overview of privacy challenges in process mining in human-centered industrial environments. 
	In \cite{tillemprivacy}, a method to secure event logs for performing process discovery by the Alpha algorithm is proposed. 
	In \cite{burattin2015toward}, the aim is to propose a solution which allows the outsourcing of process mining while ensuring confidentiality. 
	In \cite{MichaelKMBR19}, the goal is to propose a privacy-preserving system design for process mining, where a user-centered view is considered to track personal data. 
	In \cite{rafieiWA18,rafieiWA19}, a framework is proposed which provides a generic scheme for confidentiality in process mining. 
	In \cite{rafiei2019role}, the authors introduce a privacy-preserving method for discovering roles from event logs.  
	In \cite{liu2016towards}, the authors consider a cross-organizational process discovery context and share public process model fragments as safe intermediates. 
	In \cite{pretsaICPM2019}, the authors apply $k$-anonymity and $t$-closeness on event logs to preserve the privacy of \textit{resources}. 
	In \cite{MannhardtKBWM19,pripel}, the notion of \textit{differential privacy} is employed to preserve the privacy of event logs.
	In \cite{rafieitlkc}, the $\tlkc$-privacy is introduced to cope with high variability issues in event logs for applying group-based anonymization techniques.
	In \cite{batista2021uniformization}, a uniformization-based approach is proposed to preserve individuals' privacy in process mining.
	In \cite{smcProcessMining}, a secure multi-party computation solution is introduces for preserving privacy in an inter-organizational setting for \textit{process discovery}.
	In \cite{pika2020privacy}, the data privacy and utility requirements for healthcare event data are analyzed.
	In \cite{rafieippdp_arxiv}, the authors propose a privacy extension for the XES standard\footnote{https://xes-standard.org/} to manage privacy metadata.
	In \cite{riskProcessMining_short}, the authors propose a measure to evaluate the re-identification risk of event logs. Also, in \cite{rafiei_quantification}, a general privacy quantification framework, and some measures are introduced to evaluate the effectiveness of privacy preservation techniques.
	Some tools are also provided for applying the state-of-the-art privacy preservation techniques in the filed of process mining such as \textit{PPDP-PM} \cite{rafieippdpTool_arxiv}, \textit{ELPaaS} \cite{elpaas}, and \textit{Shareprom} \cite{shareprom}.

	\section{Conclusion}\label{sec:conclusions}
	In this paper, we discussed the challenges regarding directly applying traditional group-based privacy preservation techniques to event logs. We discussed the \textit{linkage attacks} and provided formal models of the possible attacks based on the different types of background knowledge. We extended the $\tlkc$-privacy for process mining to cover all the main perspectives of process mining. The data utility preservation aspect of the $\tlkc$-privacy was improved by introducing a new utility measure.
	Moreover, a new score equation was proposed to generate normalized scores for the events that need to be removed. The new equation for the score also provides \textit{privacy gain} and \textit{utility loss} coefficients that can be adjusted by users.
	Obviously, the extended version of the $\tlkc$-privacy inherits all the characteristics of the main approach. 
	Namely, it counteracts both the \textit{case linkage} and the \textit{attribute linkage} attacks. It generalizes several privacy preservation techniques including $k$-anonymity, confidence bounding, ($\alpha$,$k$)-anonymity, and $l$-diversity. 
	It also provides interpretable and tunable parameters.
	
	Similar to the main approach, we implemented four variants of the extended version with respect to the four different types of background knowledge and considering all the main perspectives. The effectiveness of different variants in different perspectives was evaluated based on real-life event logs. 
	Both \textit{data} and \textit{result} utility were analyzed to evaluate the effectiveness.
	Overall more than 1000 experiments were performed for different types of background knowledge considering different perspectives, and the results were given for a weak and a strong setting. 
	Our experiments showed that the extended $\tlkc$-privacy performs better than the previous version considering the data utility preservation aspect. However, in the event logs with the high ratio of unique traces, when the assumed type of background knowledge is very specific, e.g., \textit{relative}, the group-based privacy preservation techniques may not be able to preserve the general data utility, and this negative effect cannot be observed by only result utility analyses.

	\section*{Acknowledgment} Funded under the Excellence Strategy of the Federal Government and the L{\"a}nder. We also thank the Alexander von Humboldt (AvH) Stiftung for supporting our research.

	\bibliographystyle{splncs04}
	\bibliography{Refrences}
	
\end{document}